\documentclass[astronomy,article,accept,pdftex,moreauthors]{Definitions/mdpi} 
\firstpage{1} 
\pubvolume{1}
\issuenum{1}
\articlenumber{0}
\pubyear{2025}
\copyrightyear{2025}
 \externaleditor{    } 
\datereceived{ 24 August 2025} 
\daterevised{ 8 October 2025} 
\dateaccepted{ 20 October 2025} 
\datepublished{ } 
\hreflink{https://doi.org/} 

\newcommand{\vsini}{\ensuremath{{\upsilon}\sin i}}
\newcommand{\kms}{\,km\,s$^{-1}$}
\newcommand{\Teff}{$T_\mathrm{eff}$}
\newcommand{\logg}{$\log g$}

\Title{A Census of Chemically Peculiar Stars in Stellar Associations}

\TitleCitation{A Census of Chemically Peculiar Stars in Stellar Associations}

\Author{Lukas 
 Kueß 
 $^{1,}$*\orcidA{} and Ernst Paunzen $^{2,}$*
\orcidB{}}
 
\AuthorNames{Lukas Kueß and Ernst Paunzen}
\isAPAStyle{%
       \AuthorCitation{Kueß L., Paunzen E.}
         }{%
        \isChicagoStyle{%
        \AuthorCitation{Lukas Kueß, Ernst Paunzen}
        }{
        \AuthorCitation{Kueß, L.; Paunzen, E.}
        }
}

\address{%
$^{1}$ \quad  Department of Astrophysics, Vienna University, Türkenschanzstraße 17, A-1180 Vienna, Austria; a01405976@unet.univie.ac.at 
\\
$^{2}$ \quad  Department of Theoretical Physics and Astrophysics, Faculty of Science, Masaryk University, Kotl\'{a}\v{r}sk\'{a} 2, \mbox{611 37 Brno, Czech Republic}  
}

\corres{Correspondence:   a01405976@unet.univie.ac.at (L.K.); epaunzen@physics.muni.cz (E.P.)}

\abstract{The pre-main-sequence evolution of the chemically peculiar (CP) stars on the upper main sequence is still a vast mystery and not well understood. Our analysis of young associations and open clusters aims to find (very) young CP stars to try to put a lower boundary on the age of such objects. 
Using three catalogues of open clusters and associations, we determined membership probabilities using HDBSCAN. The hot stars from this selection were submitted to synthetic $\Delta a$ photometry, spectral, and light curve classification to determine which ones are CP stars and candidates. Subsequently, we used spectral energy distribution fitting and emission line analysis to check for possible PMS CP stars. The results were compared to the literature.
We detected 971 CP stars and candidates {in 217 clusters and associations}. A relatively large fraction, $\sim$10\% of those, show characteristics of PMS CP stars. 
This significantly expands the known list of candidate PMS CP stars, bringing us closer to solving the mystery of their origin.}


\begin{document}

\section{Introduction} \label{introduction}

Among the spectral types B, A, and F, we can find stars that show peculiarities in their spectra. These chemically peculiar (CP) stars are a well-established phenomenon since their discovery almost 130 years ago \citep{1897AnHar..28....1M}.

Classically, 
 one makes the distinction between the four types defined by \cite{1974ARA&A..12..257P} with some additions made by \cite{1996Ap&SS.237...77S}.

First,  CP1 (Am/Fm) stars show enhanced lines of iron and other metals, whereas calcium, for example, is underabundant. The fact that many of those stars are found in binary systems \citep{2020CoSka..50..570P} and thus, are rotating rather slowly (\vsini $~<~100$~\kms) leads to the so-called \textit{chemical separation}, 
 where some elements get driven outwards by radiation and others start to settle down.

The second subgroup is  CP2 or Bp/Ap or magnetic (mCP) stars. With (strong) magnetic fields up to tens of kG \citep{2021A&A...652A..31B}, they exhibit variability in their spectra and photometric time series. This is explained by the \textit{Oblique Rotator Theory} \citep{1950MNRAS.110..395S}. Currently, we are unable to explain how such strong magnetic fields can exist in those stars, given that convection is only present in their core. Nevertheless, there have been attempts made by \cite{2004Natur.431..819B}, who argue that the magnetic field is somehow frozen into the star, and by \cite{2023A&A...678A.204S}, where the authors suggest a process of the core going from radiative to convective states during the pre-main-sequence phase.

In the group of  CP3 (HgMn) stars, we find overabundances of mercury and/or manganese. Ref.
~\cite{2003A&A...397..267A} explains these peculiarities with similar processes as in  CP1 stars, suggesting a common evolution of these two objects.

The fourth group, defined by \cite{1974ARA&A..12..257P}, can be seen as an extension of  CP4 groups to hotter (early to mid-B-type) stars. They show similar characteristics in their variability \citep{1977A&AS...30...11P}. In contrast to  CP2 stars, however, the CP2 stars show primarily helium peculiarities in the form of either over- (He-strong) or underabundances (He-Weak).

In addition to \cite{1974ARA&A..12..257P}, a new group of CP stars has been discovered: the Lambda Bootis (Lam Boo) stars. These are stars with significant underabundances of almost all heavier elements. The anomalies can be explained by diffusion and mass-loss theories \citep{1986ApJ...311..326M}. Another proposed theory by \cite{1990ApJ...363..234V} proposes an interplay of accretion and diffusion, where the star passes through a metal-poor interstellar cloud. However, \cite{2022A&A...660A..98A} found little to no evidence for this mechanism.

A poorly researched question is the pre-main-sequence (PMS) evolution of those stars. There have been a few CP stars with characteristics of PMS stars reported in the literature~\citep{2024A&A...687A.176K}.

Stars in their PMS phase are characterised by accretion onto the protostar while it contracts and heats up. Commonly, two subgroups are mentioned in the literature that differ by the mass range of the protostar. Lower masses (up to $\sim$$2 \textnormal{M}_\odot$) are called T Tauri stars, while the ones above this threshold are put into the category of Herbig Ae/Be stars. Ref.
~\cite{2023SSRv..219....7B} gives an extensive overview of how those objects are characterised and how they behave. In short, one can observe them by looking for emission lines (mostly hydrogen) as a sign of the accretion phase in those stars and via  infrared excess in their spectral energy distribution~(SED).

On the other hand, we have open clusters (OCs) and stellar associations. These objects are invaluable tools for various applications in astronomy, particularly in stellar astrophysics. Astronomers use them to accurately determine stellar ages using various methods, such as isochrone fitting, kinematic studies, and the lithium depletion boundary (LDB) technique, among others. With the advent of more precise and large datasets from the $Gaia$ mission \citep{2016A&A...595A...1G, 2023A&A...674A...1G}, we now can use those astrometric measurements to accurately determine which stars belong to clusters and which do not. Recent examples working on open clusters using this dataset are \cite{2020A&A...633A..99C, 2023A&A...673A.114H}.

Stellar associations are regions of low stellar density ($\rho_\star < 0.1 \textnormal{M}_\odot pc^{-3}$, \cite{1964ARA&A...2..213B, 2023ASPC..534..129W}) which are categorised into three groups \citep{1947esa..book.....A, 1966AJ.....71..990V}:
\begin{itemize}
    \item OB associations: Mostly including hot young stars of predominantly O and B type stars (e.g., the Perseus OB1 association);
    \item T associations: Dominated by low-mass T Tauri stars in their formation stage (e.g., Chamaeleon 1);
    \item R associations: Stellar groups containing reflection nebulae  (e.g., Monocerotis R2). 

\end{itemize}

The ages of stellar associations range from a few Myr to $\sim$$10^2$~Myr \citep{2023ASPC..534..129W} and thus are well-suited to study stars right after their formation and arrival on the main sequence~\citep{2020BAAA...61C..11C}. The formation of associations is believed to have been due to one of two scenarios \citep{2020BAAA...61C..11C}: The first one states that the stars are formed where they are observed as part of a hierarchical star formation process, where over-densities in the interstellar medium can form at any scale, anywhere. The other scenario proposes that stars are formed in clusters that undergo processes such as UV radiation and stellar winds, leading to the cluster no longer being in virial equilibrium and resulting in its dispersal. 

This paper aims to look for CP stars in open clusters and stellar associations with a special interest in detecting CP candidates that are still in their PMS phase. In Section~\ref{data}, we describe the catalogues and data used, and the quality criteria employed to filter these datasets. Section \ref{astrometry} describes the transformation of the astrometry for the clustering in Section \ref{methods}. The latter also describes the process of determining membership in the aggregations, and the process of classifying stars based on their CP nature. Section \ref{pms} provides an overview of how PMS stars are detected, and the last two sections, \linebreak  Sections 
 \ref{results} and \ref{discussion}, present the results, a discussion, and an outlook on possible future work in this field.

\section{Data\label{data}} 

\subsection{List of Clusters and Associations}

Three catalogues of stellar clusters and associations were taken as a list for potential targets:

\begin{enumerate}
    \item \textit{Stars with Photometrically Young $Gaia$ Luminosities Around the Solar System} \citep{2021ApJ...917...23K};
    \item The catalogue of open clusters and moving groups by \cite{2023A&A...673A.114H};
    \item A comprehensive study about substructures in the Sco-Cen region by 
    \cite{2023A&A...677A..59R}. 

\end{enumerate}

\subsection{Used Surveys}

The analysis of the clusters, associations and stars in this list was performed using the following surveys:
\begin{itemize}
    \item $Gaia$ Data Release 3 \citep{2016A&A...595A...1G, 2023A&A...674A...1G}: With its high precision in astrometry, especially parallax (distance) and proper motion, it is ideally suited to search for open clusters and associations (e.g., \cite{2019A&A...622L..13M}). All three catalogues described above utilise this dataset. A new feature in this latest data release is the inclusion of low-resolution spectra generated by the blue and red photometers \citep{2023A&A...674A...2D,2023A&A...674A...3M}. The spectra, ranging from \mbox{300 to 1100~nm} (Figure~\ref{fig:$Gaia$_spectrum}), are great for synthetic photometry \citep{2023A&A...674A..33G} and have been proven to be effective in detecting CP stars using synthetic $\Delta a$ photometry \citep{2022A&A...667L..10P}.

\begin{figure}[H]
     
    \includegraphics[width=.9\columnwidth]{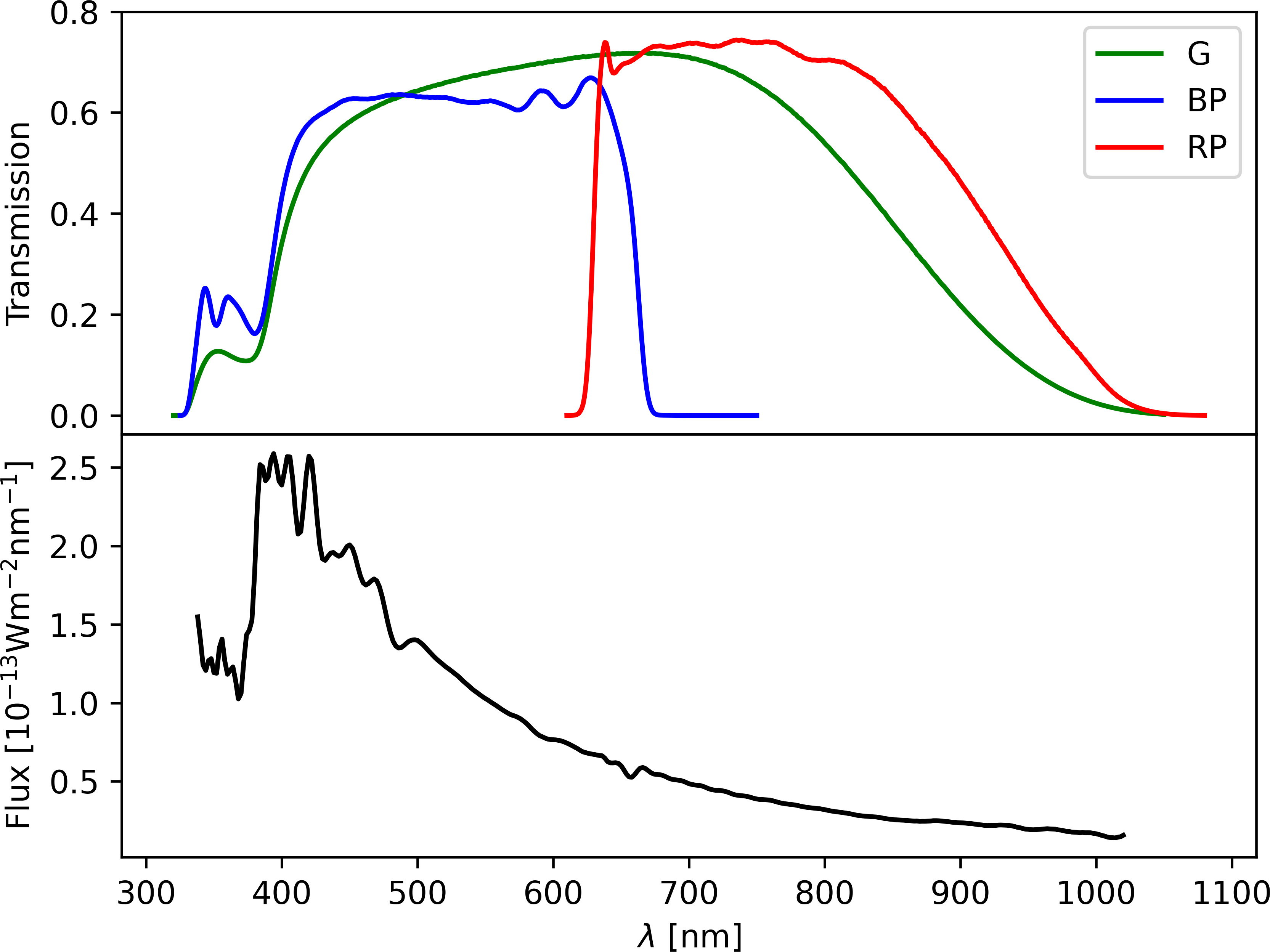}
    \caption{{{\em Upper}:} 
 $Gaia$ (E)DR3 
 passbands (\url{https://www.cosmos.esa.int/web/Gaia/edr3-passbands}, 14 January 2023) 
 {\em  Lower}:  
 Example BP/RP spectrum of an A-type star.}
    \label{fig:$Gaia$_spectrum}
\end{figure}

    \item TIC v 8.2 \cite{2021arXiv210804778P}:
    The updated source list for the Transiting Exoplanet Survey Satellite (TESS). The spacecraft was launched in 2018 and observes the sky in sectors of \mbox{$24^\circ \times 96^\circ$} with four apertures of 10\,cm \citep{2015JATIS...1a4003R}. Each sector is observed for approximately a month in two cadences: a short cadence (2 min, SC) and a long cadence (30 min, LC). As the name states, its original goal is to observe possible planet-hosting stars. However, the time series data created by the telescope is an excellent source for studying variable stars. The instrument has one drawback: the pixel size is relatively large ($21" \times 21"$), making the possibility of blending in more crowded fields a big problem (e.g., \cite{2022A&A...666A.142S}). 
    \item Data Release 9 (\url{http://www.lamost.org/dr9/}, accessed in April 2024) 
 of the Large Sky Area Multi-Object Fiber
    Spectroscopic Telescope (LAMOST, \cite{2012RAA....12..723Z,2012RAA....12.1197C}): Data from this Multi-Object spectrograph based in China have a spectral resolution of $R \sim$ 1000--1500 
and thus are well-suited for spectral classification. This has led to the discovery of numerous CP stars in the past  \citep{2020A&A...640A..40H,2022ApJS..259...63S}. However, due to its limited movement, it can observe the sky only in the northern part, i.e., between $\delta \in [-10^\circ, +90^\circ]$. 
\end{itemize}

\subsection{Data Retrieval and Quality Assessments} \label{data_retrieval}

The data for the clusters and associations were downloaded in a region around the astrometric parameters of each object. This means all stars within the parallax range $[\varpi_{min}, \varpi_{max}]$ of the cluster, an angular radius $r$ around its core and within the boundaries of the proper motions $[\mu_{\alpha,min}, \mu_{\alpha,max}]$ and $[\mu_{\delta,min}, \mu_{\delta,max}]$, were retrieved from the $Gaia$ archive using a python script, and after a recommended zero-point correction on the parallax \citep{2021A&A...649A...4L}, they were stored in a \texttt{.csv} file for each region on the sky. Generally, a few key quality criteria were taken during this step.

First, for the lists from \cite{2021ApJ...917...23K,2023A&A...677A..59R}, a quality cut on the parallax and its error was taken, leaving only sources with $\varpi/\sigma_\varpi >10$. In the case of \cite{2023A&A...673A.114H}, this value was lowered to $\varpi/\sigma_\varpi >5$ in order to include fainter sources with less accurate astrometry due to the large distances of clusters in this catalogue. Additionally, Ref. 
 \cite{2023A&A...673A.114H} gives a parameter called ``astrometric signal-to-noise ratio'' of which they argue that a value above five means that the cluster is likely to be real. Accordingly, all clusters below this threshold were removed, along with the 133 globular clusters included in their catalogue.

For all the datasets, we took a cut in the re-normalised unit weight error (RUWE), removing poor astrometric solutions and binaries, leaving only stars with $\textnormal{RUWE}<1.4$ in the final dataset \citep{LL:LL-124}.

After these astrometric quality cuts, some additional quality criteria on the photometric measurements from $Gaia$ had to be taken into account. It is recommended to remove all sources with $G_{BP}>20.3$ mag due to a systematic brightening effect in those faint regions~\cite{2021A&A...649A...3R}. The same paper also recommends a cut on the so-called ``corrected flux-excess factor'' $C^*$ that is defined as 
\begin{equation}
    C^* = C - f(G_{BP}-G_{RP})
\end{equation}
where $C$ is the original flux excess factor defined as the combined flux in the blue and red photometers divided by the total flux in $G$:
\begin{equation}
    C = \frac{F{G_{BP}}+F_{G_{RP}}}{F_G}
\end{equation}
\indent \textls[-25]{See \cite{2021A&A...649A...3R} for details on this correction.
We used a cut of $3\sigma$ on the corrected flux excess factor~$C^*$.}

\section{Astrometric Transformations\label{astrometry}} 

After all the quality criteria described above were imposed, some astrometric transformations had to be taken into account for the planned clustering on the datasets.

At first, the parallax and galactic coordinates $l$, $b$ were transformed into 3D Euclidean distances $X$, $Y$, $Z$ using a standard transformation:
\begin{equation}
\begin{pmatrix}
X \\
Y \\
Z
\end{pmatrix}
=
R
\begin{pmatrix}
\cos l \cos b \\
\sin l \cos b \\
\sin b
\end{pmatrix}.
\end{equation}
where $R$ is simply calculated by taking the inverse of the parallax in mas to get to a distance of pc:
\begin{equation}
    R = \frac{1000}{\varpi}
\end{equation}

The next step was to transform the proper motions in the International Celestial Reference System (ICRS) frame $\mu_{\alpha_\star}( = \mu_\alpha \cdot \cos \delta)$, $\mu_\delta$ to galactic proper motions $\mu_{l_\star}$, $\mu_b$. This was done utilising the vectors and matrices described by the $Gaia$ DR3 documentation (\url{https://gea.esac.esa.int/archive/documentation/GDR3/Data_processing/chap_cu3ast/sec_cu3ast_intro/ssec_cu3ast_intro_tansforms.html}, 28 February 2023). 

Finally, the proper motions in the galactic frame were transformed into transverse velocities (in km/s) via
\begin{equation}
    \begin{pmatrix}
        v_{T_{l_\star}}\\
        v_{T_b}
    \end{pmatrix}
     = 4.74\cdot R
    \begin{pmatrix}
        \mu_{l_\star}\\
        \mu_b
    \end{pmatrix}
\end{equation}

This finally resulted in a five-dimensional dataset ($X,Y,Z,v_{T_{l_\star}},v_{T_b}$) that we then used for the subsequent membership analysis of the clusters and associations.

\section{Methods \label{methods}}

\subsection{Clustering}

Determining the membership of stars in open clusters and associations is not always a straightforward task, and depending on the data and methods used, one can obtain vastly different results in membership numbers and probabilities. Commonly used methods are Gaussian Mixture Models (GMMs, e.g., \cite{Agarwal_2021}), Unsupervised Photometric Membership Assignment in Stellar Clusters (UPMASK, \cite{2014A&A...561A..57K}) or, more recently, Significant Mode Analysis (SiGMA, \cite{2023A&A...677A..59R}). Other methods try to use a combination of photometry, astrometry, and spectroscopy to determine which stars are members of young associations (e.g., \cite{2007ApJS..173..104L}).

To determine which stars of our list were likely members of open clusters and associations, we used the HDBSCAN algorithm \citep{McInnes2017} that was also used by \cite{2021ApJ...917...23K,2023A&A...673A.114H}.

HDBSCAN is a clustering algorithm with several possible hyperparameters in its implementation. The most important ones are the following:

\begin{itemize}
    \item \texttt{min\_cluster\_size}: This 
 is somewhat self-explanatory; it describes how many points should be at a minimum in a cluster. Larger values will result in smaller substructures that get lumped together in a bigger cluster.
    \item \texttt{min\_samples}: This describes how dense a cluster should be. Larger values will result in only the densest cores being considered to be clusters.
    \item \texttt{cluster\_selection\_method}: There are two choices for this parameter:
        \begin{itemize}
            \item \texttt{eom}:
            This stands for ``excess of mass'' and searches for over-densities in the n-dimensional parameter space, looking for clusters in accordance with the other parameters. It generally favours larger and fewer clusters.
            \item \texttt{leaf}:
            This lets the user recover finer and smaller clusters in the dataset.
        \end{itemize}
    \item \texttt{metric}: This chooses the distance metric that the algorithm uses to calculate the distance between the points in the input set. Commonly used choices are \texttt{euclidean}, \texttt{manhattan} or \texttt{malahanobis}.
\end{itemize}

Of course, HDBSCAN is not perfect and has some shortcomings. According to \cite{2023A&A...673A.114H}, it tends to be a bit overconfident, meaning that the algorithm assigns more data points to the cluster than there actually are. However, they used an all-sky approach, whereas this work only clusters the stars in the regions around the clusters and associations. This should result in at least the cores of the aggregations to be recovered. 

\begin{figure}[H]
     
    \includegraphics[width=.93\textwidth]{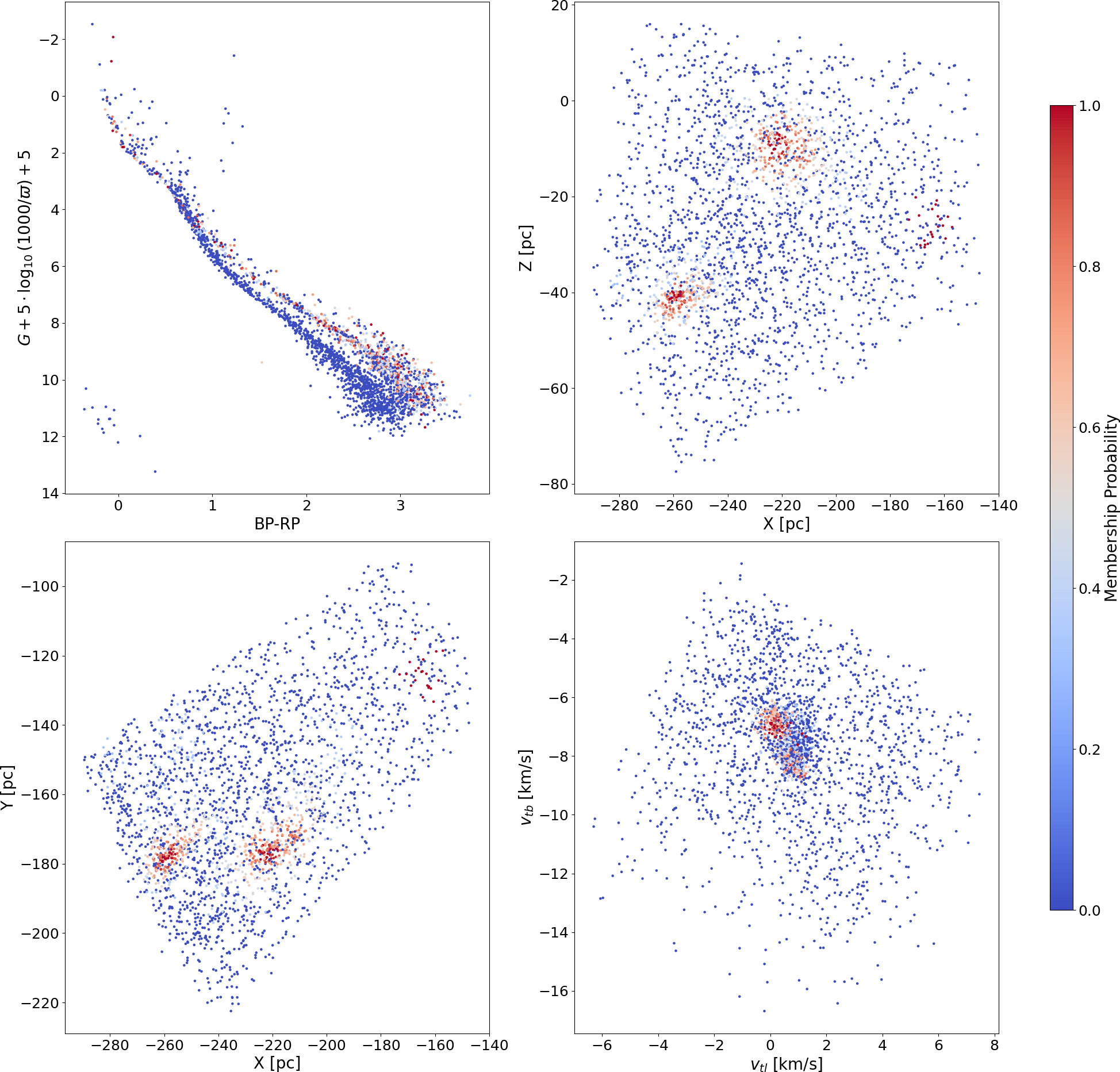}
    \caption{\textls[-15]{Result 
 of the clustering on the Monocerotis south-west region, called as such by \cite{2021ApJ...917...23K}. The stars are colour-coded by membership probability. Three of the sub-plots show astrometric properties ($XYZ$-distances and transversal velocities). The fourth one on the upper left shows a colour--magnitude diagram (CMD) of the region as a visual check if the clustering gives a reasonable~result.}}
    \label{fig:clustering}
\end{figure}

We randomly selected 30\% of the data in each cluster region to look for the best hyperparameter combination of the values seen in Table \ref{tab:hdb_params}. The different combinations were evaluated using a silhouette score, from which the highest, i.e., the best parameter combination, was chosen for the final clustering on the set. An example of the clustering result can be seen in Figure \ref{fig:clustering}.
In total, the clustering resulted in a list of roughly 462.5~thousand stars that have a high ($p>0.5$) probability of being members of the clusters and associations in question after removing stars that are below the main sequence (i.e., White Dwarfs).
From the 430 thousand stars in 2022 clusters and moving groups that resulted from the clustering in and around the catalogue of~\cite{2023A&A...673A.114H}, around 200 k also belong to the initial catalogue {of~\cite{2023A&A...673A.114H}}. This is a recovery rate of 15\%, a result of the difference in the clustering method: Ref. 
 \cite{2023A&A...673A.114H} used an all-sky method, whereas we used the regions around the individual clusters and aggregations. For the catalogue from \cite{2021ApJ...917...23K} we recovered a similar fraction (13\% or $\sim$12,500~stars in 26 associations) due to the same reasons of a different approach to the clustering. This discrepancy in membership, of course, may affect the number/fraction of CP stars detected in these groups.

\unskip
\begin{table}[H]
     
    \caption{Hyperparameter range searched.\label{tab:hdb_params}}
    
    \begin{tabularx}{\textwidth}{CC}
     
    \toprule
        \textbf{Parameter} & \textbf{Range of Values} \\
        \midrule
         
         \texttt{min\_cluster\_size} & 5, 50] 
 in steps of 5\\
         \texttt{min\_samples} & [3, 5, 10, 20, 30, 40]
\\
         \texttt{cluster\_selection\_method} & [\texttt{eom}, \texttt{leaf}]\\
         \texttt{metric} & [\texttt{euclidean}, \texttt{manhattan}]\\
         \texttt{algorithm} & [\texttt{best}]\\
         \bottomrule
    \end{tabularx}
\end{table}

\subsection{Extinction Correction and Final Target Selection}

\begin{figure}[H]
     
    \includegraphics[width=.71\columnwidth]{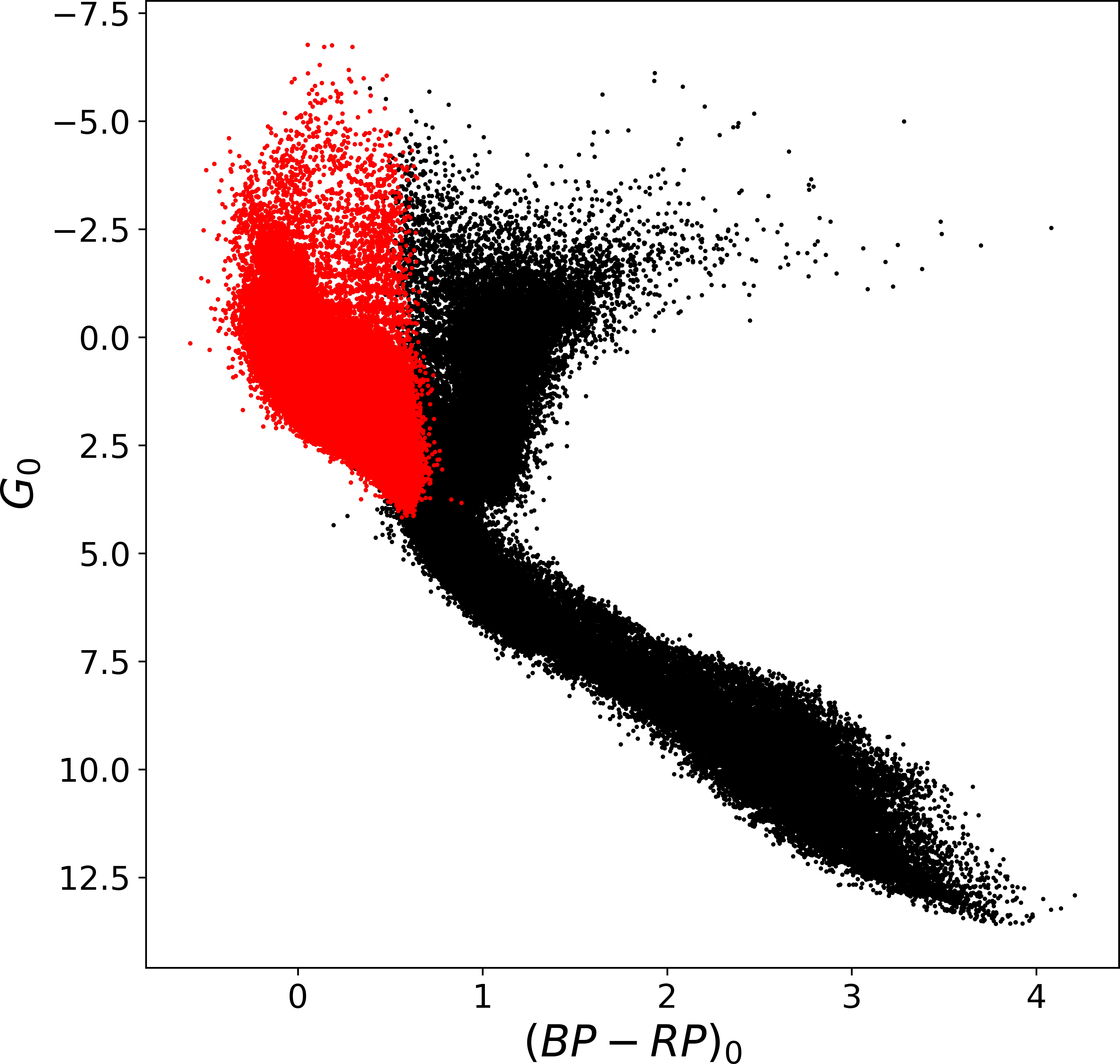}
    \caption{Extinction 
 corrected Colour--Absolute Magnitude Diagram of the clustered stars in the StarHorse21 catalogue. The red stars 
 denote the stars hotter than $T_{eff}>6500$~K.}
    \label{fig:final_selec}
\end{figure}

This list, however, is still not sufficient to sort out all the stars that are not in question of being chemically peculiar. Since we wanted to look at stars with $T_\mathrm{eff}>6500$~K (or hotter than spectral type around mid-F), the cooler stars had to be removed from the list. Although we also wanted to look at PMS CP stars, we opted for this boundary, since there has been no evidence for PMS CP stars in T Tauri stars that we could find in the literature.  {There might be other classes of peculiar stars in the lower temperature region, predominately barium stars. These, however, are not in the focus of the current work.} All previously discovered PMS CP stars belong to the class of Herbig Be/Ae stars; see \cite{2025Astro...4...15K} and references therein.
For this purpose, we took the StarHorse21 catalogue from \cite{2022A&A...658A..91A}. The authors of this paper compiled measurements from $Gaia$ EDR3, the Two-Micron All-Sky Survey (2MASS,~\cite{2006AJ....131.1163S}), the Panoramic Survey Telescope and Rapid Response System (Pan-STARRS1, \cite{2016arXiv161205560C}), and the Wide-Field Infrared Survey Explorer All-Sky Release (AllWISE, \cite{2013wise.rept....1C}) to derive extinction values $A_V$ at $\lambda = 5420$~\AA, and stellar parameters such as \Teff, \logg, $M/\textnormal{M}_\odot$, and $XYZ$-positions with respect to the galactic center. Approximately 333 thousand stars of the clustered sample are included in this catalogue. Out of those, around a third, 105,270
~stars, have a temperature higher than the threshold as mentioned above. See Figure~\ref{fig:final_selec} for the CMD of this final target list. We want to emphasise the point that the choice of another catalogue of stellar parameters might result in a different number of stars selected. See \cite{2025Astro...4...15K} for a discussion of the influence of different catalogues on this matter. 

\subsection{$\Delta a$ Photometry}\label{syn_phot}

\begin{figure}[H]
     
    \includegraphics[width=0.6\columnwidth]{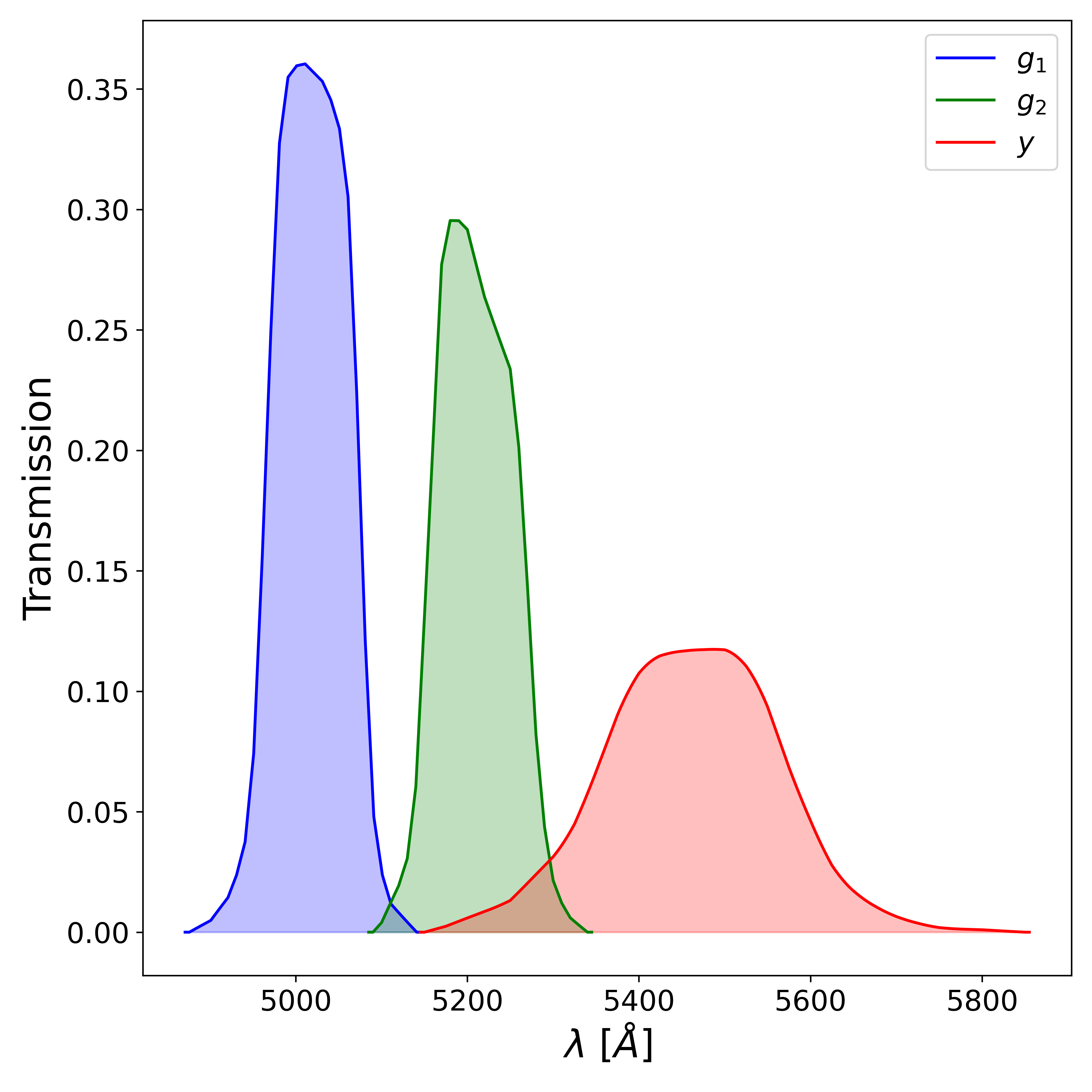}
    \caption{Filter curves used for $\Delta a$ photometry.}
    \label{fig:filters}
\end{figure}

The first method we used for detecting possible CP stars was $\Delta a$ photometry. This powerful method, developed by \cite{1976A&A....51..223M}, makes use of the flux depression in stellar spectra around $\lambda 5200$. First reported by \cite{1969ApJ...157L..59K} in the spectrum of HD 221568, this depression has been proven to indicate the presence of a strong magnetic field in mCP stars. The most critical element affected here is iron, although there are also contributions of silicon, chromium, and others \citep{2007A&A...469.1083K}.
The method itself is relatively simple. One takes three filters, $g_1$ to the left (shorter $\lambda$) of the depression, $g_2$ right at the depression, and the Strömgren $y$ filter (see Figure \ref{fig:filters}) at the right (longer $\lambda$) of the flux depression. Then the magnitudes in each filter are measured or calculated synthetically. These magnitudes can then be converted into an index $a$ that is defined as 
\begin{equation}
    a = g_2-\frac{g_1+y}{2}
\end{equation}

This value $a$ is then compared to the value of a (seemingly) normal star that has a value $a_0$ to create  $\Delta a$:
\begin{equation}
    \Delta a = a - a_0(g_1-y)
\end{equation}

The colour term $(g_1-y)$ corrects for temperature differences in the stars, and the function $a_0(g_1-y)$ is then called the "normality line". All stars that are a certain distance (typically, one uses 95\% or $3\sigma$ confidence intervals) above this line are considered to be magnetic stars with a significant flux depression.

Before the synthetic photometry, the list of the clustered stars was matched to the catalogue of the $Gaia$ BP/RP (XP in short) spectra via the $Gaia$ DR3 IDs, of which 64 thousand stars had a spectrum. For each of those, the signal-to-noise ratio (S/N) was calculated using the quick way, described in \cite{2023A&A...674A...2D}. Basically, since the spectra are downloaded as a set of coefficients for each passband, one can estimate the S/N by the $L_2$ norm of the coefficient vector divided by the $L_2$ norm of the vector of errors on the coefficients: 
\begin{equation}
    S/N_{XP} = \frac{\| \texttt{xp\_coefficients} \|_2}{\| \texttt{xp\_coefficient\_errors} \|_2}
\end{equation}

The final estimation was done by just taking the mean of the S/N in each passband. All stars with an $S/N>100$ were considered for the subsequent analysis. This left 26,814
~spectra to be analysed using synthetic $\Delta a$ photometry.

The photometry was done in the same way as described in \cite{2022A&A...667L..10P}. At first, the spectra that initially came in flux units of W
 m$^{-2}$nm$^{-1}$ were normalised at a common wavelength of $\lambda = 4020$~\AA. Additionally, they were interpolated using a third-order polynomial technique to match the resolution of the used filters, which have $\Delta \lambda = 1$~\AA.

Actual astronomical magnitudes in the AB system \citep{1983ApJ...266..713O}, for example, for a given filter curve $f(\lambda)$ and a spectrum $s(\lambda)$ are calculated using the following quantity \citep{2012PASP..124..140B}:
\begin{equation}
    m_f = -2.5 \cdot \log \frac{\int f(\lambda) s(\lambda)\lambda d\lambda}{\int f(\lambda)(c/\lambda) d\lambda} -56.10
\end{equation}

However, since the interest here only lies in the difference of (instrumental) magnitudes, one can take the much more uncomplicated approach where the filter and the spectrum are just multiplied and the values are then summed up:
\begin{equation}
    m_f = \sum_i f(\lambda_i) s(\lambda_i)
    \label{eq:synth_phot}
\end{equation}

After we calculated the synthetic magnitudes this way, the normality line of the form $a_0 = c_1 + c_2 \cdot (g_1-y)$, i.e., the coefficients $c_1, c_2$ needed to be found. For this purpose, we used the Python 3.10 package 
 \texttt{PyMC} \citep{AbrilPla2023PyMC}. This uses the theorem of Bayes
\begin{equation}
    P(\hat{\theta}\vert \hat{X},\hat{y}) = \frac{P(\hat{y}\vert\hat{X},\hat{\theta})P(\hat{\theta})}{P(\hat{y}\vert \hat{X})}
\end{equation}
with the input vector $\hat{X} = (g_1-y)$ and $\hat{y} = a$ to infer the vector of coefficients $\hat{\theta}$ via a Monte Carlo process. The resulting normality line (see also Figure \ref{fig:delta_a}) had the following form:

\begin{figure}[H]
     
    \includegraphics[width = \columnwidth]{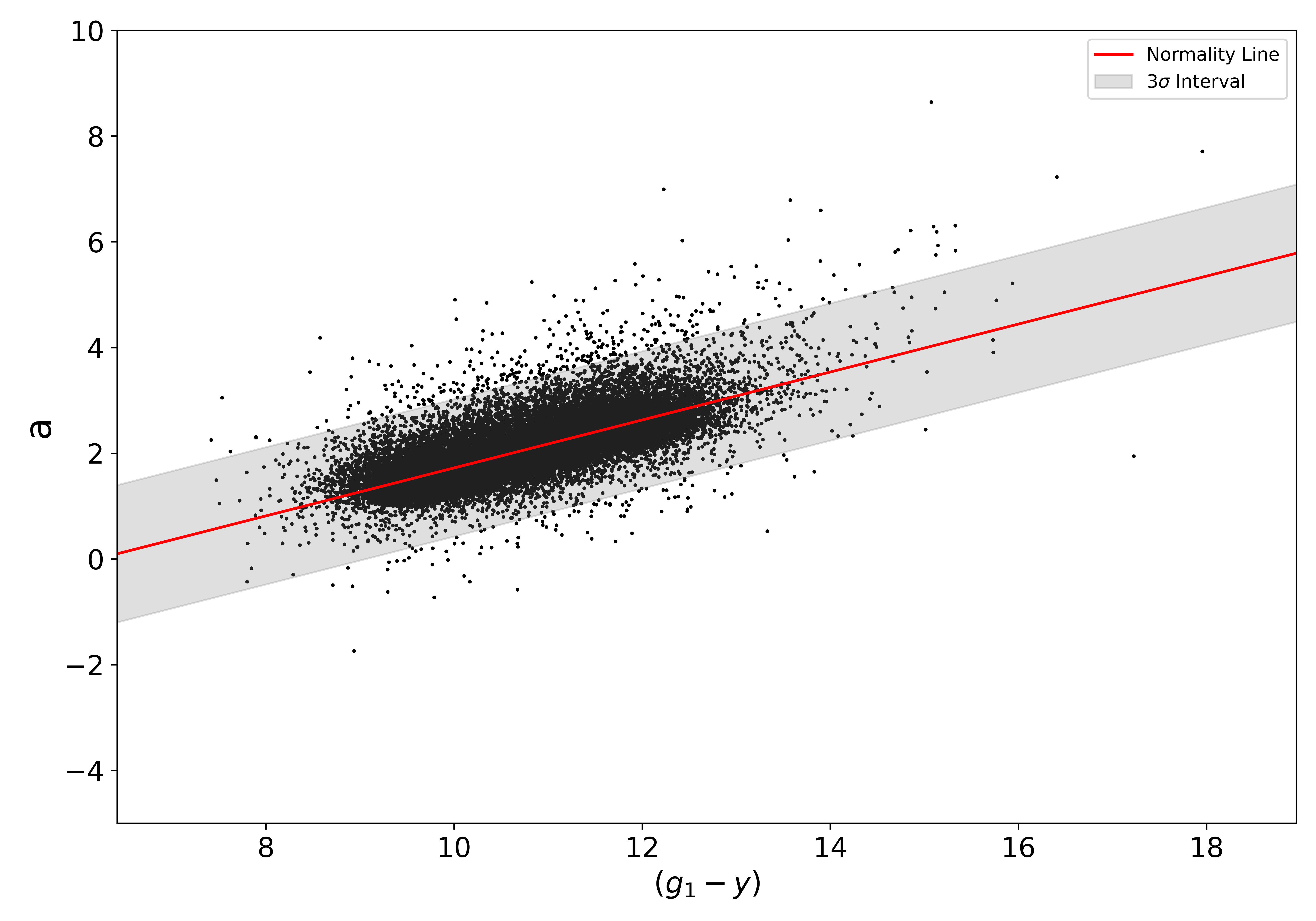}
    \caption{Result 
 of the synthetic $\Delta a$ photometry. See Section \ref{syn_phot} for details.}
    \label{fig:delta_a}
\end{figure}

\begin{equation}
    a_0 = 0.454(3)(g_1-y) - 2.815(29)
\end{equation}

Additionally, \texttt{PyMC} calculated the standard deviation $\sigma$ of this fit line, and it gave a value of $\sigma \approx 0.433$. All stars with a distance of more than $3\sigma \approx 1.298$ above $a_0$ are considered to be candidate mCP stars, which amounted to 341 individual stars.

\subsection{Spectral Classification} 

Another method to find CP stars is the classification of stellar spectra. The $Gaia$ DR3 IDs of the 105 thousand candidate stars on the upper main sequence were submitted to the ninth data release of LAMOST (\url{http://www.lamost.org/dr9/}, accessed in April 2024),  
which contained 4143~spectra of this set. Removing all spectra with S/N lower than 50 in the $g$-band and, in the case of stars with multiple spectra, keeping the ones with higher S/N left us with 1440~spectra to classify. Those were classified using the MKCLASS code \citep{2014AJ....147...80G}. This program takes an input spectrum and compares it to a standard library to derive the final spectral type after some iterations. The results are comparable to those made by humans, with only minor deviations in the temperature and luminosity classes. It is also well-suited to classify CP stars as done by \cite{2020A&A...640A..40H}, for example.

For the classifications, we used three standard libraries:
\begin{itemize}
    \item \textit{libnor36}; 
    \item \textit{libr18};
    \item \textit{libr18\_225}.
\end{itemize}

Out of the 1440 spectra, 608 individual stars showed CP characteristics in their spectra for at least one of these libraries. This high fraction of roughly 42\% shows that either an unusual amount of stars in young open clusters and associations are CP stars or that some stars have been misclassified by one or more standard libraries of MKCLASS. Especially, \textit{libr18\_225} has a high recovery rate of CP1 stars, and thus, one has to be careful with the classifications.

Of course, one has to be cautious with the spectra of lower quality, although they should still result in reasonable estimations of the spectral type.

\subsection{Light Curve Analysis}

\begin{figure}[H]
     
    \includegraphics[width=\textwidth]{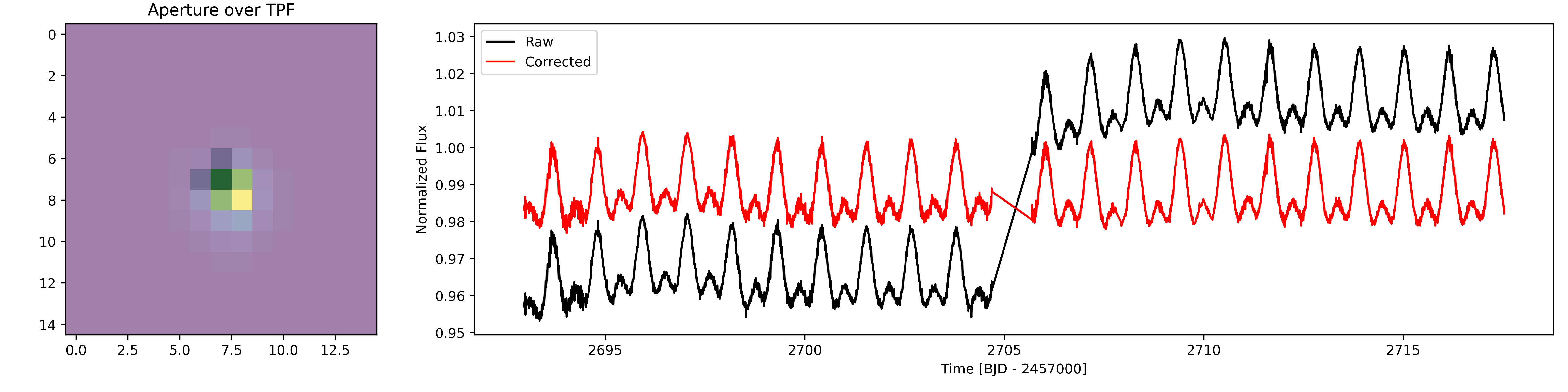}
    \caption{Process 
 of extracting 
 the lightcurves for a circular mask (Section \ref{methods}).   {\em Left}: Target 
 pixel file with aperture.  {\em  Right}: Raw (black) and corrected (red) normalised flux. The corrected flux values have been shifted down by 1.5 per cent for clarity.}
    \label{fig:eleanor}
\end{figure}

A volume-limited subset of the stars with \Teff$>6500$~K was also checked for light curves from TESS. This was limited in volume to the list of \cite{2021ApJ...917...23K} because of the large pixel size of TESS (Section \ref{data}) and to minimise blending as much as possible. The light curves of 1022 stars were downloaded using the Python package \texttt{eleanor} \citep{2019PASP..131i4502F}. For the process of retrieving the time series data, an aperture with a radius of one pixel was used (Figure~\ref{fig:eleanor}). The resulting data were cut by 3$\sigma$ to remove outliers and then converted from flux differences to magnitude differences via Pogson's equation
\begin{equation}
    \Delta mag = -2.5 \cdot \log_{10}\left(\frac{F}{F_0}\right)
\end{equation}
where $F_0$ is the mean flux of the observations.

After this process, we used the astropy \citep{astropy:2013, astropy:2018, astropy:2022} implementation of the Lomb--Scargle periodogram \citep{1976Ap&SS..39..447L, 1982ApJ...263..835S} that follows the algorithm described in \cite{2009A&A...496..577Z}. Additionally, one can calculate the false-alarm probability (FAP) using the methods from \cite{2008MNRAS.385.1279B} to check if the found periods are significant and thus correct. We used a value of $\log(\textnormal{FAP})\leq-3$ for a period to be significant. This process was done in two frequency regions:
\begin{itemize}
    \item $0.001\,\textnormal{c/d}\,\leq\,f\,\leq5\,\textnormal{c/d}$;
    \item $f \, \geq 5 \,\textnormal{c/d}$.
\end{itemize}

We then classified the light curves based on their frequency spectrum and phase-folded light curve (Figures 
 \ref{fig:dsct} and \ref{fig:acv}). Generally, we followed the criteria given in \cite{2022A&A...666A.142S} while analysing the light curves with the types of variability taken from the Variable Star Index (VSX, \cite{2006SASS...25...47W}) and the notation from the General Catalogue of Variable Stars (GCVS, \cite{2017ARep...61...80S}). The most prominent variability classes in our sample are pulsators such as Gamma Doradus (GDOR) or Delta Scuti stars (DSCT), Rotating stars (ROT) and their subgroup of $\alpha$ Canum Venaticorum (ACV) stars, and stars with variability of unknown origin or unclear signal (VAR). A general scheme we used can be seen in Table \ref{tab:classification}.

\begin{table}[H]
         \caption{Basic classification scheme for the visual classification of variability types. This is a  short overview of the most prominent classes in the sample, a more detailed description can be found in~\cite{2022A&A...666A.142S}.     \label{tab:classification}}

\begin{adjustwidth}{-\extralength}{0cm}

    \begin{tabularx}{\fulllength}{ccC}
     
    \toprule
    \textbf{Type} & \textbf{Features in Light Curve} & \textbf{Features in Periodogram} \\
     
    \midrule
     GDOR & (ir)regular, beating, sharp curves &  two or more peaks $<5$ c/d\\
     DSCT & beating, short periods  & two or more peaks $>5$ c/d\\
     ROT & sinusoid variation, repeating pattern with stable features & Peaks at harmonics of rotational frequency\\
     ACV & rotation with ``double waves'' due to chemical spots & Same as ROT but clear CP2 stars from LC and/or spectrum\\
     VAR & variability present, no clear features & one or more peaks with unknown origin\\
     \bottomrule
    \end{tabularx}

\end{adjustwidth}
\end{table} 

Classification of light curves presents several challenges. These range from unclear signals in the periodograms, poor data, ambiguity in classification due to a lack of more detailed information about the star in question, and blending effects. Accordingly, there is a non-negligible chance that light curves will be classified incorrectly. See e.g., \cite{2022A&A...666A.142S} for a more detailed discussion on that matter.

\begin{figure}[H]
     
    \includegraphics[width=.96\columnwidth]{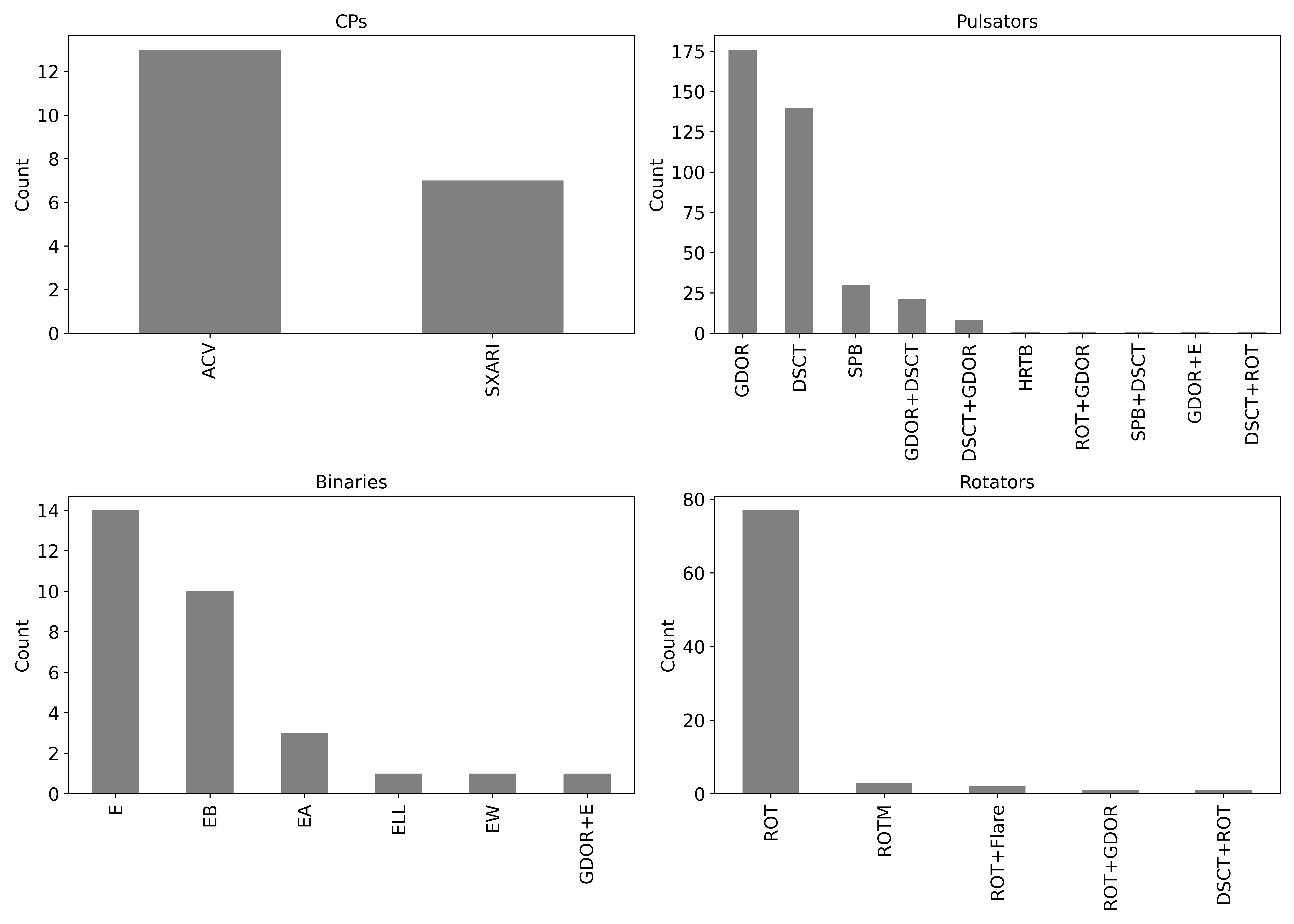}
    \caption{Result of the classification from the TESS light curves. Note that some mixed types appear in more than one subplot.}
    \label{fig:vars_classes}
\end{figure}

However, 512 (50\%) out of the 1022 light curves showed apparent variability and could be classified. The result can be seen in Figure 
 \ref{fig:vars_classes}. A total of 22 stars had light curves with attributes from CP stars, i.e., a so-called ``double wave'' from the Oblique Rotator \citep{1950MNRAS.110..395S}. The classes in question are ACV (after $\alpha²$ CVn, also  {Figure} \ref{fig:acv}) and SXARI (after SX Arietis), which are essentially hotter analogues of the ACV variables.

Altogether, 971 individual CP stars and candidates in 217 open clusters and associations were detected from at least one of those three methods.

{\begin{figure}[H]
     
    \includegraphics[width=.96\columnwidth]{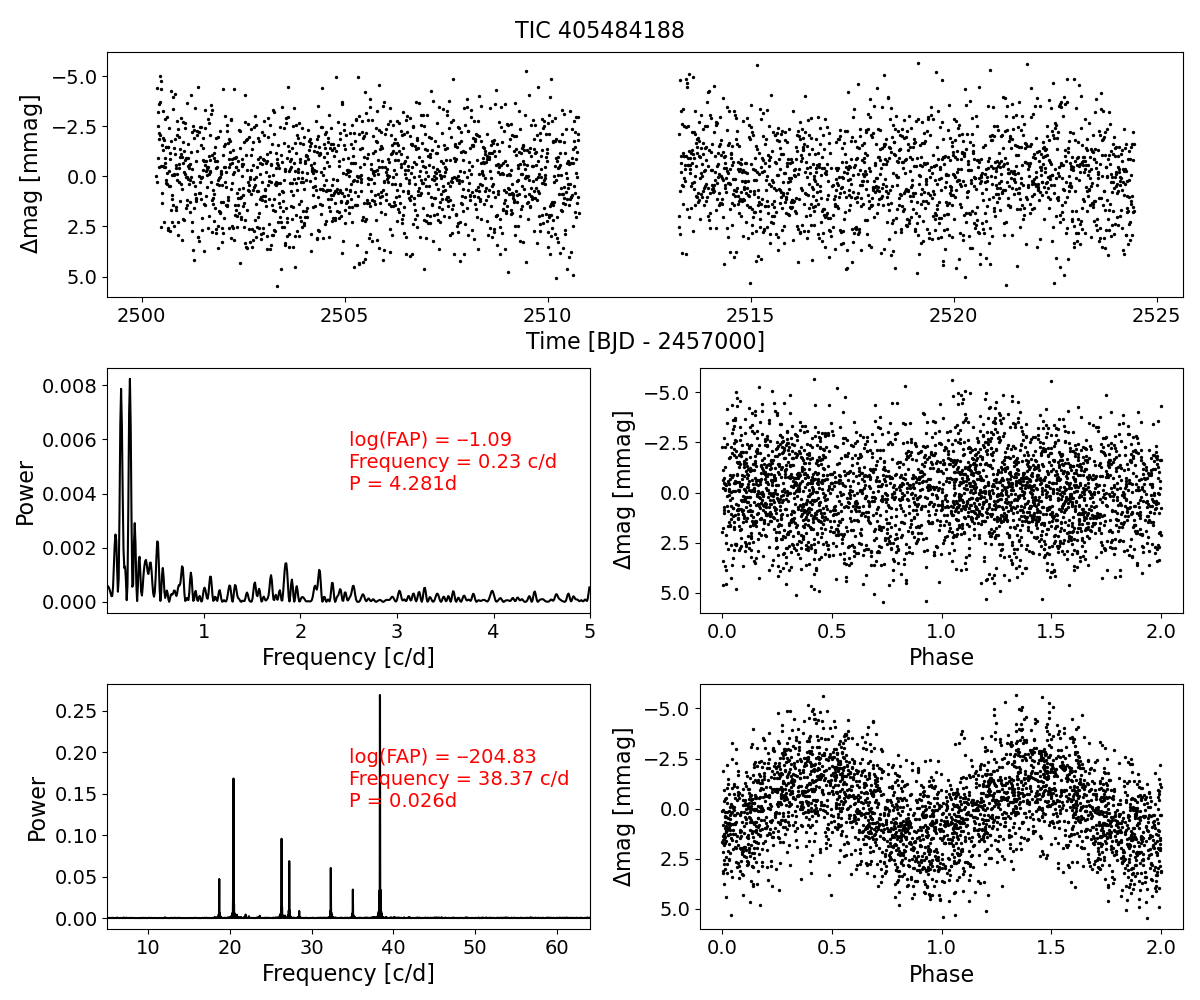}
    \caption{\textit{Upper}: Final 
 light  
 curve of the DSCT variable TIC 405484188 as observed with TESS. \textit{Middle-left}: Lomb--Scargle
    periodogram in the region up to 5 c/d. \textit{Middle-right}: Phased light curve of this frequency region. The small bump in the minima (=the double wave)
    is clearly visible. \textit{Lower-left}: Periodogram for frequencies above 5 c/d, note
    the comparatively small power of the peaks.  \textit{Lower-right}: Phase-folded light
    curve at the highest peak above 5 c/d.~~One can clearly~see~~multiple}
     \label{fig:dsct}
\end{figure}

\vspace{-12pt}
 
  \captionof*{figure}{\textls[-15]{individual peaks in this region, which are also typically observed in pulsating stars, such as DSCT stars.
    In the periodograms, the frequency of the highest power, the corresponding
    period, and the FAP are given in~red. }}}

\begin{figure}[H]
     
    \includegraphics[width=.96\columnwidth]{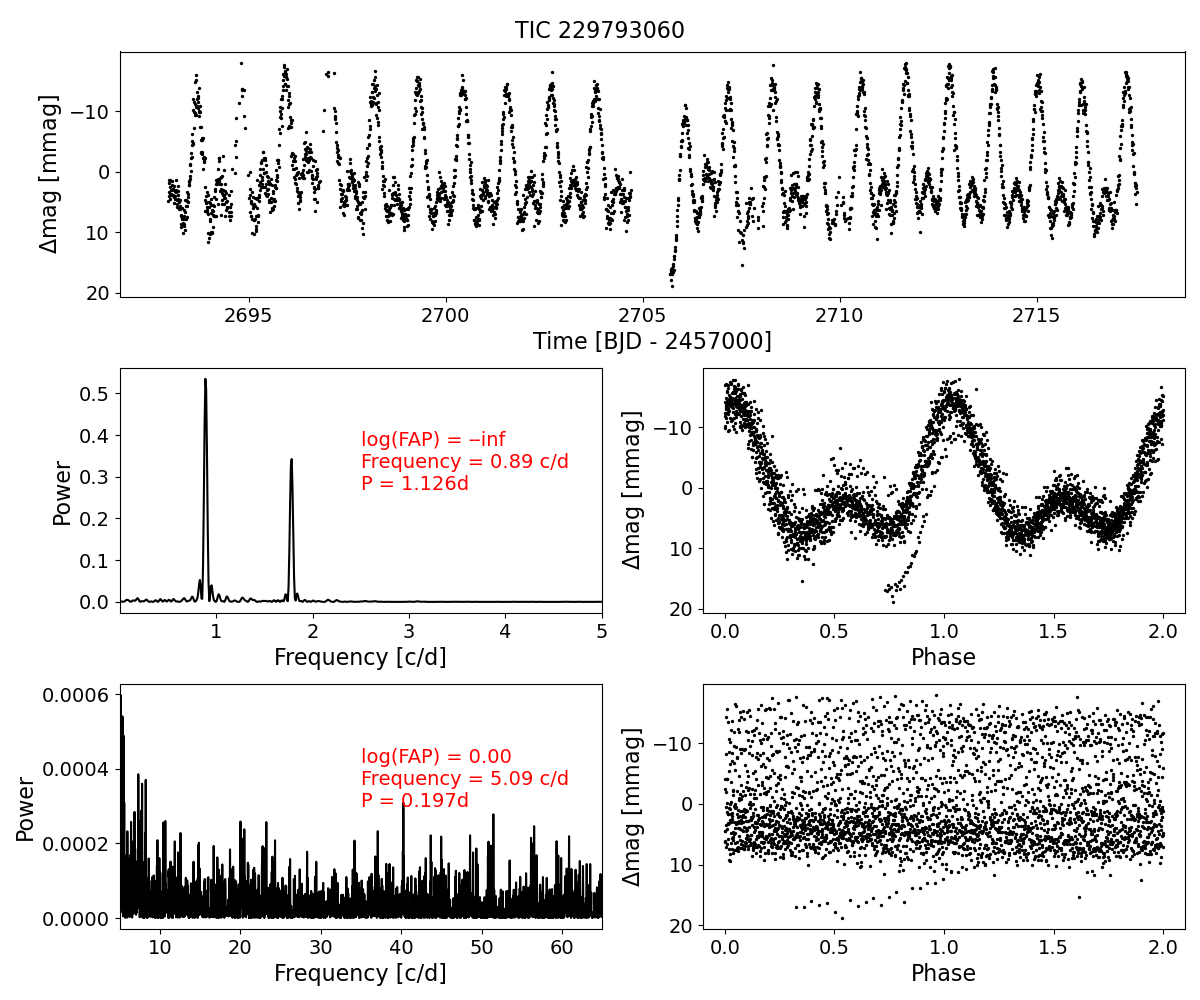}
    \caption{Same  
   as  
 Figure \ref{fig:dsct}, but for the ACV variable TIC 229793060 (EE Dra/HD 177410). Note the different scales on the y-axis in the periodogram plots.}
    \label{fig:acv}
\end{figure}

\unskip

\section{Pre-Main-Sequence Stars} \label{pms}

Finally, from all the detected CP stars, we wanted to assess how many of those are still in their PMS or accretion phase. To achieve this, one can employ two primary methods: fitting the spectral energy distribution (SED) or searching for emission in the Balmer lines of the spectrum. 

\subsection{Spectral Energy Distributions}

The SEDs of the stars in question were created using the Virtual Observatory SED Analyser (VOSA, 
\url{http://svo2.cab.inta-csic.es/theory/vosa/}, accessed on 2 October 2024, \cite{2008A&A...492..277B}). This tool essentially searches for available distances, extinction values, and photometry. Then it fits an SED to the data and also allows one to determine accurate parameters such as radius, mass, and luminosity of the star in question, depending on the accuracy of the input parameters.

\begin{figure}[H]
     
    \includegraphics[width=.9\columnwidth]{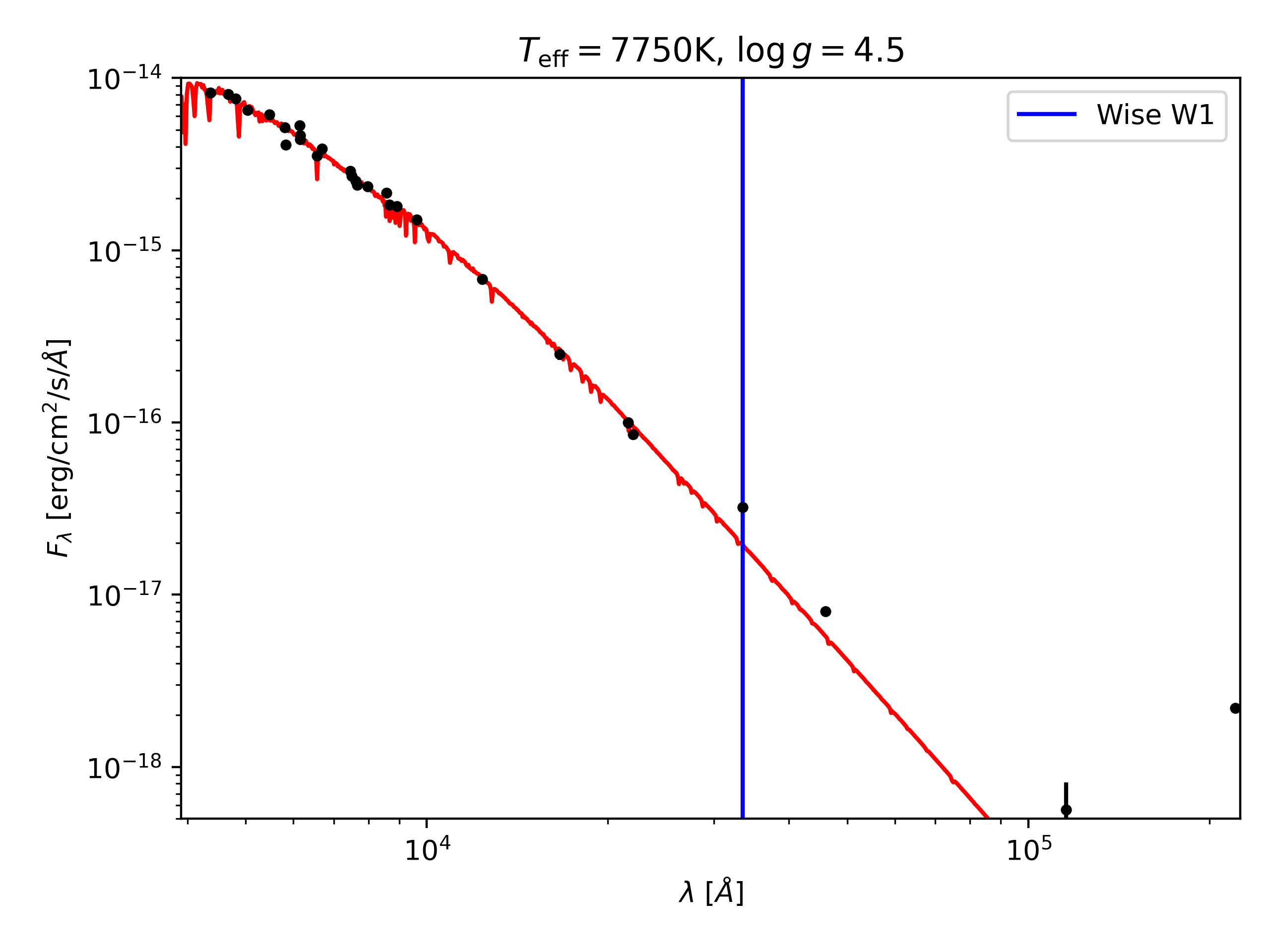}\vspace{-6pt}
    \caption{SED of a PMS CP star. The black dots and error bars denote the photometry calculated from VOSA, and the red spectrum is a model with the given parameters \Teff\, and \logg. The model was taken from \cite{2003IAUS..210P.A20C}.}
    \label{fig:excess}
\end{figure}

We took the coordinates of the 971 stars in question and submitted them to VOSA. One could also include other parameters, such as magnitudes or fluxes in one or more filters, distances, and extinction values ($A_V$). However, to avoid inhomogeneities in one or more of these values, we stuck with the coordinates.
An SED with infrared excess is seen in Figure~\ref{fig:excess}. Out of the whole sample of 971 CP stars and candidates, 92 show IR excess (Table~\ref{tab: ir_excess_table}).
\vspace{-6pt}

\subsection{Emission Lines}

\begin{figure}[H]
     
    \includegraphics[width=.9\columnwidth]{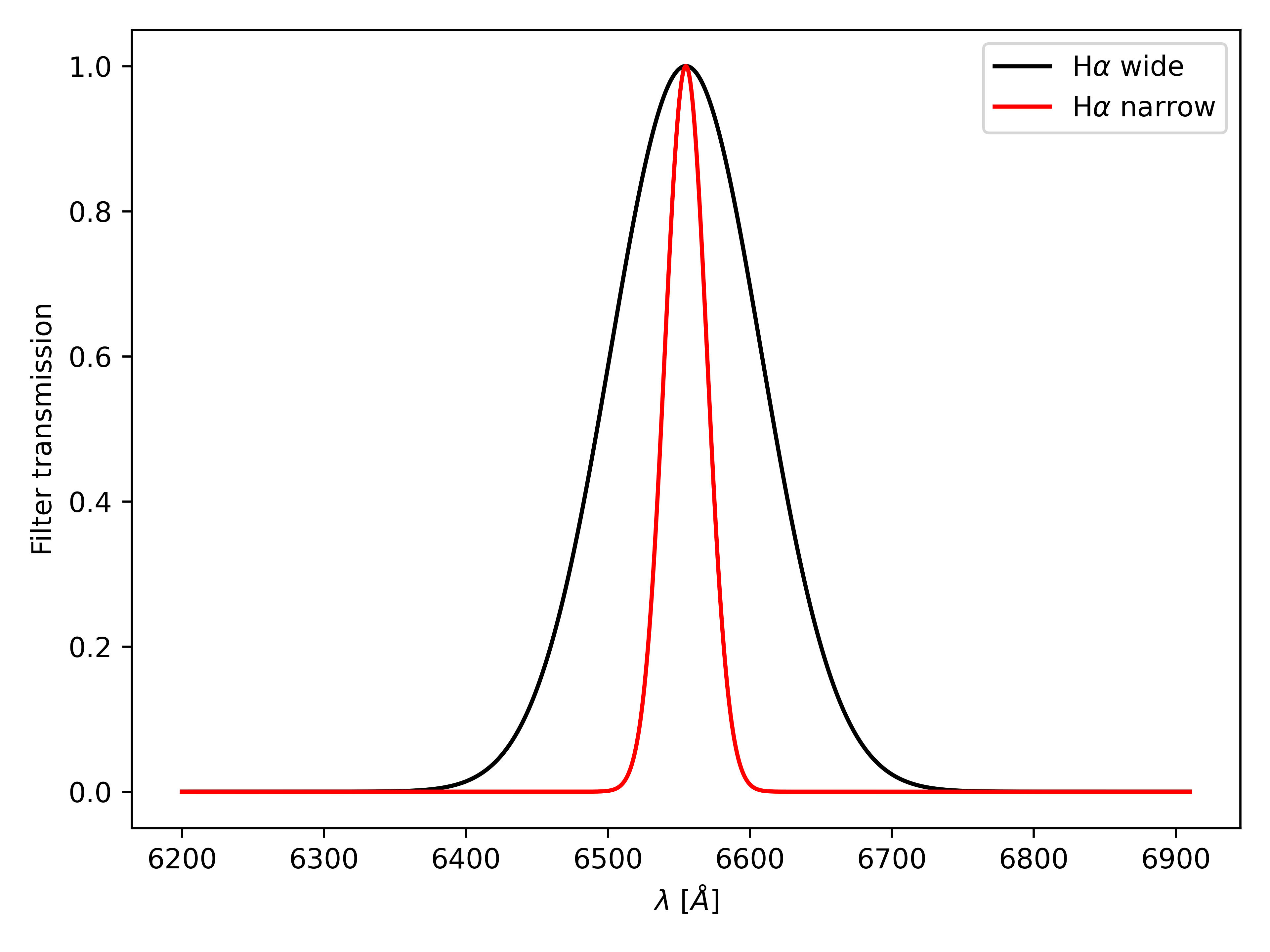}
    \caption{H$\alpha$ filters used in the emission line photometry.}
    \label{fig:alpha_filters}
\end{figure}

The other method we used to detect stars in their PMS phase is the detection of emission lines of hydrogen, especially H$\alpha$.
Ref.  
\cite{halpha2024} made an approach similar to the $\Delta a$ photometry to detect emission lines. They took one wide and one narrow synthetic (Gaussian) filter with a Full-Width at Half-Maximum (FWHM) of 125~\AA\, and 35~\AA, respectively. The filters are both centred at 6555~\AA\, (Figure \ref{fig:alpha_filters}). The filters were then folded with the 
 $Gaia$ XP spectra of a sample of known Herbig stars, and then compared the synthetic magnitudes to gain an~index:
\begin{equation}
    H\alpha = \frac{H\alpha_{\textnormal{wide}}}{H\alpha_{\textnormal{narrow}}}
\end{equation}

This is similar to the $\beta$-index from Stömgren photometry \citep{1966ARA&A...4..433S} used to determine effective temperatures of stars.

If this value is below a certain threshold, one can assume the star exhibits emission in H$\alpha$ and is, thus, likely a PMS object.

We took this approach and used the LAMOST spectra to determine the spectral types of the stars. They were treated in the same way as the XP spectra for the synthetic photometry, i.e., normalised at $\lambda 4020$ and interpolated to a resolution of 1~\AA. We also had to modify the filters, shifting the central wavelength to 6564~\AA\, to get the best results. Then, the magnitudes were calculated in the same way as in Section \ref{syn_phot}.

\begin{figure}[H]
     
    \includegraphics[width=.91\columnwidth]{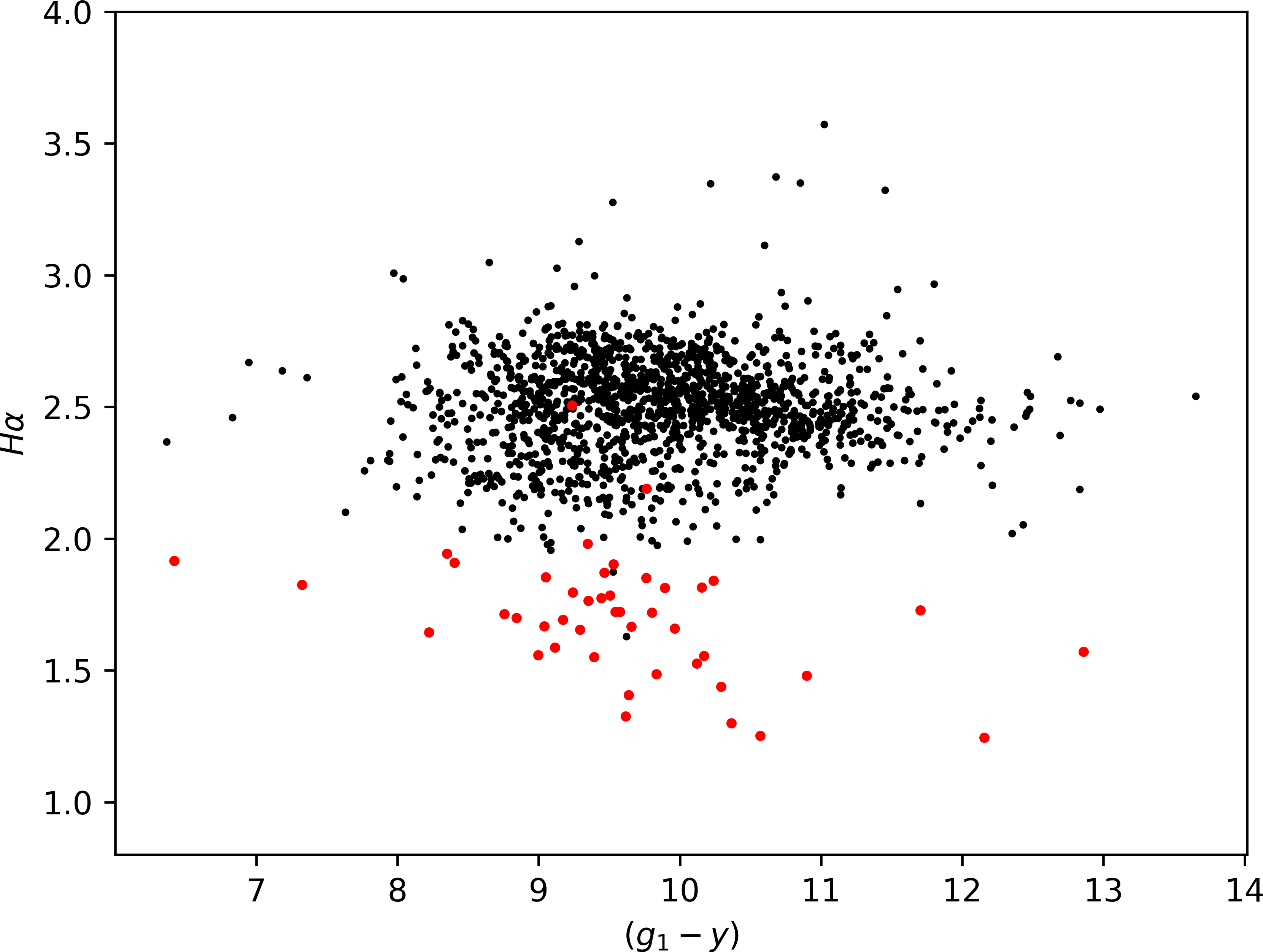}
    \caption{Result of the synthetic photometry from the H$\alpha$ filters on the LAMOST spectra. The black dots represent all stars and the red ones show the emission. One can easily see that almost every star with $H\alpha \lesssim 2$ shows emission.}
    \label{fig:emission_lines}
\end{figure}

Additionally, the spectra were checked with the Python package \texttt{specutils} \citep{2019ascl.soft02012A} to search for emission lines in the region around the H$\alpha$ line. The combined product of these two processes can be seen in Figure \ref{fig:emission_lines}. From all the stars with emission lines, twelve are also (candidate) CP stars (Table \ref{tab:emission_stars}).

\unskip

\section{Comparison with Literature}

\textls[-15]{To verify whether we had identified any previously known CP stars, we also performed cross-matching with catalogues from the literature.
The catalogues in question~are as follows:}

\begin{enumerate}
    \item Ref. 
 \cite{2009A&A...498..961R}:
        A collection of CP stars from the literature. The authors compiled a list of 8205 stars that have been previously flagged as CP stars. However, no assessment of the correctness of the classifications was done, so some of them have been confirmed, while others have been proven to be ``normal'' stars. A total of 156 stars in the target list of  stars hotter than 6500K are contained in this catalogue, of which 9 could be detected in this work.
    \item Ref. \cite{2019ApJS..242...13Q}:
    The authors of this paper searched 200,000 
 stars from LAMOST DR5 for CP features using a random forest classifier together with an additional manual classification. They publish a list of 9372 CP1 and 1132 CP2 stars. Subsequently, they investigated on the spatial distribution of the stars. This catalogue contains 27 objects, which could also be recovered using my analysis.
    \item Ref. \cite{2020A&A...640A..40H}:
    This catalogue of 1002 CP stars was conceived using spectra from LAMOST DR4 and a modified version of MKCLASS. Additionally, they checked for the flux depression using $\Delta a$ photometry. The authors checked for the evolutionary status on the main sequence, mass, and spatial distribution. Four stars from the list of possible targets in this work can be found in their catalogue, half of which I could detect.
    \item Ref. \cite{2020MNRAS.496..832C}: 
    This catalogue uses spectra from the Apache Point Observatory Galactic Evolution Experiment (APOGEE, \cite{2017AJ....154...94M}) to look for HgMn (CP3) stars. A total of 260 new CP3 stars were catalogued by their work, of which I could recover seven.
    \item Ref. \cite{2022ApJS..259...63S}:
    Here, the authors used an XGBOOST algorithm and manual classification on LAMOST DR8. They arrived at a list of 17,986
 CP1 and 2708 CP2 stars and included astrophysical parameters (\Teff, \logg, $[Fe/H]$) of the stars. In total, 15 out of the 17 stars that are also in my target list could be rediscovered by my methods.
    \item Ref. \cite{2024ApJS..272...43Y}:
    The authors used MKCLASS and an ensemble regression model with spectra from LAMOST DR10. Out of the 27,543 
  CP 
 stars and candidates, they list $\sim$11,614~new Am stars and candidates as well as 4978 new Ap stars and candidates. They also include stellar parameters (\Teff, \logg, $[Fe/H]$) in their catalogue. I could recover 59~stars from their list.
\end{enumerate}

Whenever possible, we used the $Gaia$ DR3 IDs to avoid mismatching, and in all the other cases, a search radius of 1 arcsecond was used to minimise the risk of false identification.

We note the low recovery rates when cross-matching to the catalogues. {The reasons for this are not clear. Effects arising from far distances can likely be ruled out since all the catalogues and surveys mentioned above also include far and faint sources. A possible explanation may be due to the low coverage of the densest parts of the galactic plane by large spectroscopic surveys and the TESS mission. We are also aware that a part of our CP stars and candidates may have been misclassified by either the automatic or manual classification of the spectra and light curves in our sample or some of the catalogues using machine learning algorithms to classify their spectra.}

\section{Results\label{results}} 

\subsection{$\Delta a$ Photometry}

Looking at the flux depression in stars with $Gaia$ XP spectra, 341 could be detected using $\Delta a$ photometry. Three of the detected stars additionally have a spectral type from LAMOST, all being CP2 stars (Table \ref{tab:spec_delta}). See Table \ref{tab:delta_detec} for the first entries of the list of spectral classifications. The entire list will be available online. A colour--magnitude diagram (CMD) can be seen in Figure \ref{fig:delta_cmd}.

\begin{table}[H]
\footnotesize
    \caption{Stars that have a positive $\Delta a$ value and a spectral type from  LAMOST.
    \label{tab:spec_delta} 
}

\begin{adjustwidth}{-\extralength}{0cm}

   \begin{tabularx}{\fulllength}{cccccccC}
 \toprule
        \textbf{\boldmath{\textit{$Gaia$}}  DR3} & \textbf{LAMOST} & \boldmath{$\alpha$} & \boldmath{$\delta$} &  {{\texttt{\textbf{libnor36}}}} &  {{\texttt{\textbf{libr18}}}} &  {{\texttt{\textbf{libr18\_225}}}} &  {\boldmath{$\Delta a$}} \\
         
        \midrule
        253428535333057920 
 & J042756.29+442830.8 & 66.985 & 44.475 & kB9.5hA1mA2  Si & kB9.5hA2mA5  SrSi & kA0hF0mA6  SiEu & 1.519 \\
        253436747311206528 & J042906.36+442429.2 & 67.277 & 44.408 & A0 II-III  Si & A0 II-III  Si & F1 V Fe-5.9 & 2.014 \\
        3344712726726459520 & J061909.56+143120.1 & 94.790 & 14.522 & kA1hA2mA5  Si & A0 III-IV  SrSi & kA1hA8mA8  SiEu & 1.357 \\
        \bottomrule
    \end{tabularx}

\end{adjustwidth}
\end{table}

\subsection{Spectral Classification}

\begin{figure}[H]
     
    \includegraphics[width=\textwidth]{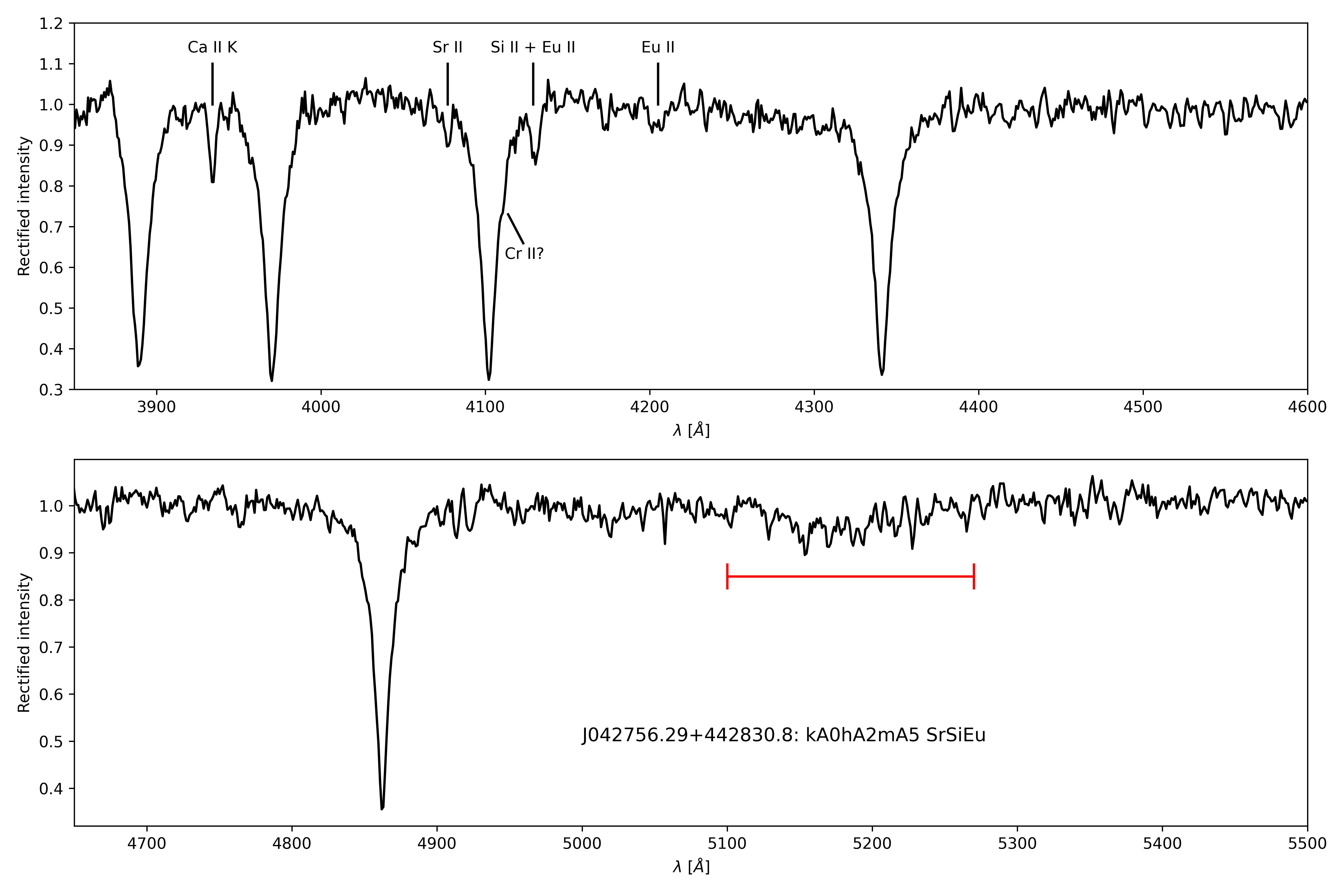}
    \caption{Spectrum  
 of the mCP star $Gaia$ DR3  253428535333057920/LAMOST
    J042756.29+442830.8. 
 The spectral type indicated is the result of a manual check I made for a few randomly selected spectra after the classification. A few spectral types of interest are marked in the upper panel, and the approximate extent of the flux depression discussed earlier.}
    \label{fig:mcp_spec}
\end{figure}

From the LAMOST spectra, we detected 608 CP stars in three standard libraries. One example spectrum of a newly detected CP2 star is shown in Figure \ref{fig:mcp_spec}. Additionally, a few new Lambda Bootis stars were detected. An example is in Figure \ref{fig:lam_boo_ex}. Note that not all spectra were of good quality, so there is a chance that some of them have been classified incorrectly. A CMD of these stars can be seen in Figure \ref{fig:cmd_spec}. The beginning of this catalogue is presented in Table \ref{tab:spec_detections}; the complete list will also be available online.

\subsection{Photometric Time Series}

Out of the approximately 1000 stars that had light curves available from TESS, 20~showed variability connected to CP stars, namely ACV or SXARI variability. The periods range from approximately one-third of a day to five days, which aligns with previous findings on these variables. A CMD of the variable CP stars can be seen in Figure \ref{fig:cmd_lcs} and the list with basic astrometry, periods, and FAP can be found in Table \ref{tab:lc_detec}.

\subsection{PMS Stars}

From the analysis of the CP candidates in terms of their evolutionary status, we found that twelve stars show emission lines in H$\alpha$. Additionally, 92 stars showed infrared excess in their SED, another sign of an early evolutionary status. See Table \ref{tab: ir_excess_table} for the stars and the start of the IR excess as well as Table \ref{tab:emission_stars} for the stars with H$\alpha$ emission.

\subsection{Binarity}

Binarity might play a significant role in the CP phenomenon, as already described in Section \ref{introduction}.
Basically, the CP stars rotate much more slowly than the ``normal'' stars in the same spectral region. 
To reduce the angular momentum, binary systems are most suitable. The general binarity rate of all
B- and A-type stars is about 50\% \citep{2013ARA&A..51..269D}.
For the CP stars, there is no recent study of the binarity rate available. However, looking into the
corresponding literature, we find a value of 50\% \citep{2020CoSka..50..570P,2021A&A...645A..34P}, 
even for the CP1
stars \citep{2023MmSAI..94b..50C}.
The $Gaia$ DR3 catalogue contains 1.5 
 $\times$ 10$^9$ sources with a full, single-star
astrometric solution, but potentially also contains a large number of binary systems. The presence
of an unresolved companion can affect the single-star solution \citep{2020MNRAS.495..321P} and thus
the parallax and proper motions, most important to derive the membership probability to a stellar
association, for example. Undetected binary systems can be traced by an elevated RUWE value
\citep{2020MNRAS.496.1922B,2024A&A...688A...1C}.
As described in Section \ref{data_retrieval},  we took a cut in the RUWE parameter of 1.4, respectively.
This should already exclude the binary systems for which the $Gaia$ single-star solution is not
valid. Our sample comprises 20 objects with a flag of multiple source identifiers. Those are good
candidates for being binary systems. But it can also mean that there are other close members of the 
stellar association. Finally, we also checked for known wide and astrometric binaries discovered by $Gaia$ 
\citep{2020ApJS..247...66H,2021MNRAS.506.2269E}. In total, we found thirteen matches or slightly above 1\%.
We can, therefore, conclude that our sample consists mainly of single stars.

\section{Summary and Outlook\label{discussion}} 

Using different methods ($\Delta a$ photometry, spectral, and light curve classification), we detected 971 CP stars in 217 open clusters and associations; most of those CP candidates have not been named in the literature before. The membership probabilities of the stars in the respective stellar aggregation were determined using HDBSCAN. Additionally, we used SED-fitting and emission line photometry to determine which CP stars are likely to be in their PMS phase.

We found that 12 CP stars show emission in H$\alpha$, whereas 92 CP stars show infrared excess. Note that the excess is relatively weak in most cases, suggesting that the stars are already in the late PMS evolution stage and close to  the Zero-Age Main Sequence (ZAMS).

The stars with emission lines all belong to the category of CP2 (mCP) stars, agreeing with the suggestion that the magnetic field is at least partially responsible for the occurrence of CP stars \citep{2024A&A...687A.176K}. However, this is still a tiny sample and a much more detailed investigation is needed to confirm or disprove these ideas.

Nevertheless, the present list of about 100 new PMS CP candidates vastly expands the list of known PMS sources with CP properties, which will be a stepping stone for future research on the formation of these objects, particularly the origin and early evolution of the magnetic fields in stars on the upper main sequence and the onset of the diffusion processes in these objects.
A recent study by \cite{2025MNRAS.536...72F} already discusses the occurence of CP stars in open clusters and also made an analysis of the evolutionary status. They discuss a decrease in CP stars in the evolution from the ZAMS to the terminal-age main sequence (TAMS). We did not analyse the present sample in this way due to the relatively large number of unconfirmed CP candidates. This, however, is surely an opportunity for future research with higher quality of spectral data from future telescope such as the Extremely Large Telescope which is currently under construction. 
\vspace{6pt}

\authorcontributions{C
onceptualization, E.P., methodology: L.K, formal analysis, L.K.,
writing–original draft preparation: L.K., writing–review and editing, L.K. and E.P.; visualization, L.K. All the authors have read and agreed to the published version of the manuscript.}

\funding{This research received no external funding} 




\dataavailability{The original contributions presented in this study are included in the article. Further inquiries can be directed to the corresponding authors.} 

\acknowledgments{This work has made use of data from the European Space Agency (ESA) mission
{\it Gaia} (\url{https://www.cosmos.esa.int/gaia} 
 processed by the {\it Gaia}
Data Processing and Analysis Consortium (DPAC,
\url{https://www.cosmos.esa.int/web/gaia/dpac/consortium}). Funding for the DPAC
has been provided by national institutions, in particular the institutions
participating in the {\it Gaia} Multilateral Agreement.
This paper includes data collected with the TESS mission, obtained from the MAST data archive at the Space Telescope Science Institute (STScI). Funding for the TESS mission is provided by the NASA Explorer Program. STScI is operated by the Association of Universities for Research in Astronomy, Inc., under NASA contract NAS 5–26555.
Guoshoujing Telescope (the Large Sky Area Multi-Object Fiber Spectroscopic Telescope LAMOST) is a National Major Scientific Project built by the Chinese Academy of Sciences. Funding for the project has been provided by the National Development and Reform Commission. LAMOST is operated and managed by the National Astronomical Observatories, Chinese Academy of Sciences.}

\conflictsofinterest{The authors declare no conflict of interest.}

\appendixtitles{yes} 
\appendixstart
\appendix
\section[\appendixname~\thesection]{Colour--Magnitude Diagrams}


\begin{figure}[H]
     
    \includegraphics[width=\textwidth]{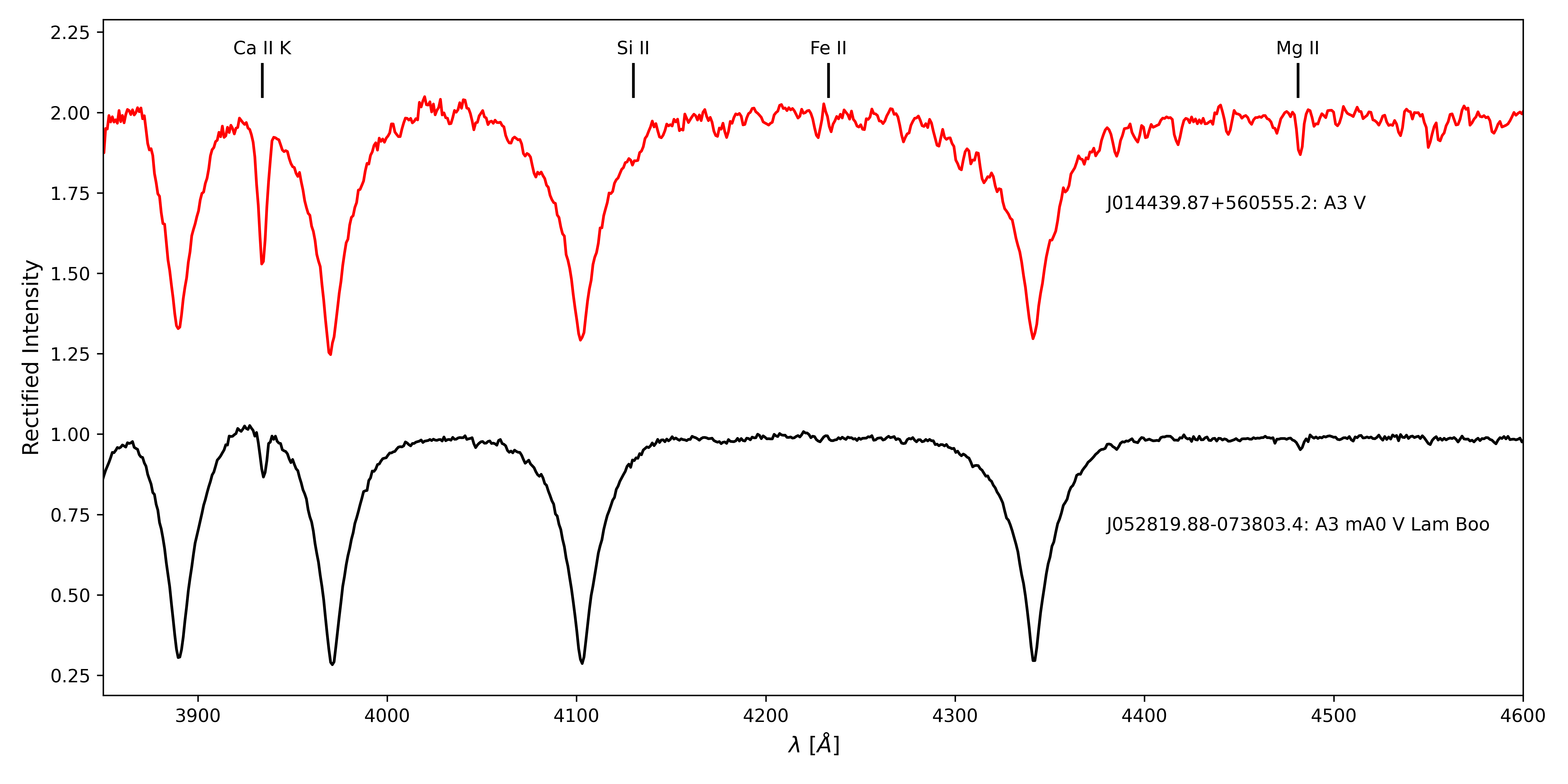}
    \caption{One 
 newly 
 detected Lambda Bootis star. As one can easily see, there are (almost) no metal lines present apart from the (very) weak Ca II K line and the Mg II line at $\lambda 4481$.}
    \label{fig:lam_boo_ex}
\end{figure} 
\unskip
\begin{figure}[H]
     
    \includegraphics[width=.9\columnwidth]{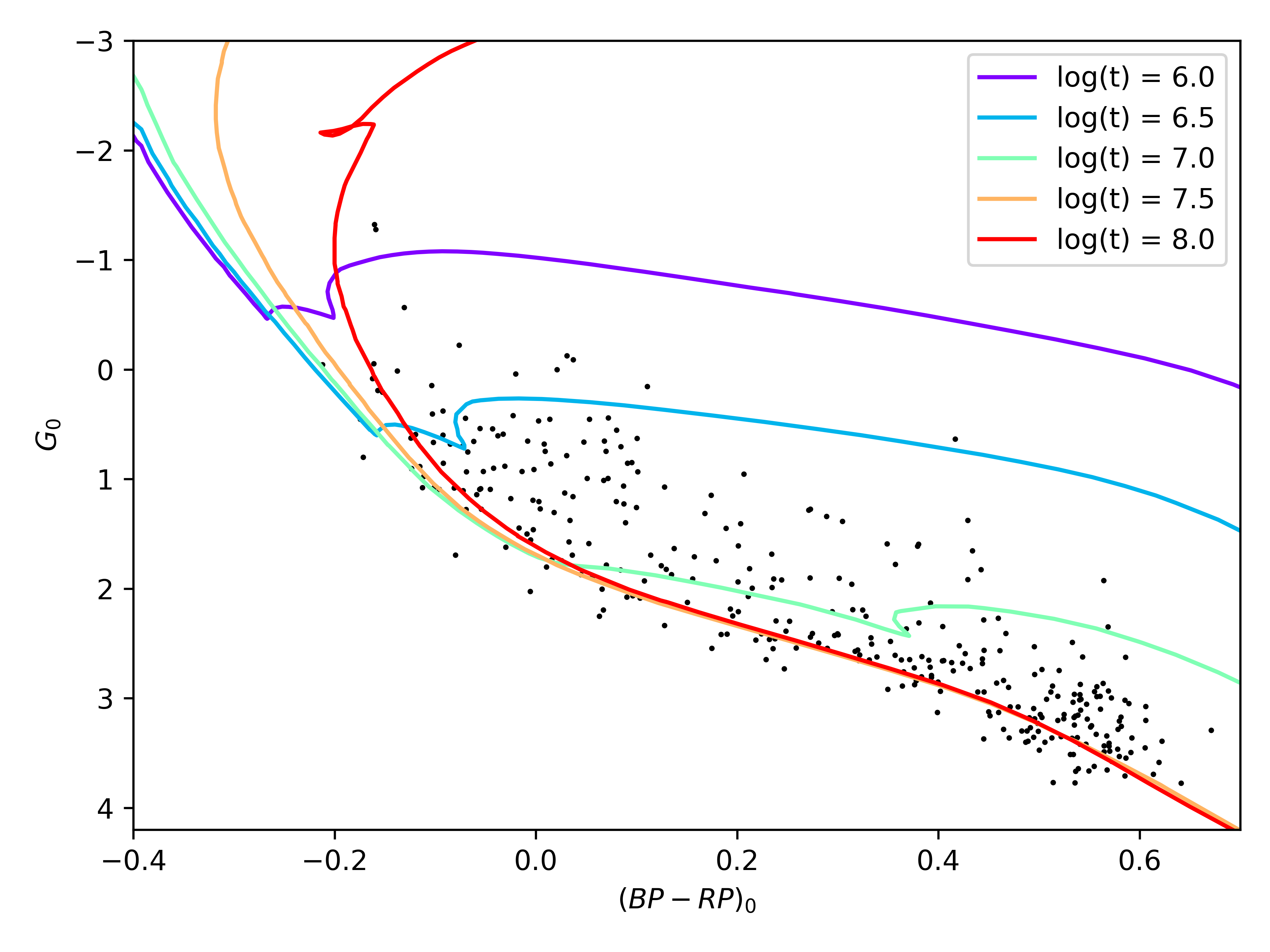}
    \caption{CMD 
 of the stars that were detected using synthetic photometry. The isochrones were taken from \cite{2012MNRAS.427..127B}.}
    \label{fig:delta_cmd}
\end{figure}
\unskip
\begin{figure}[H]
     
    \includegraphics[width=.9\columnwidth]{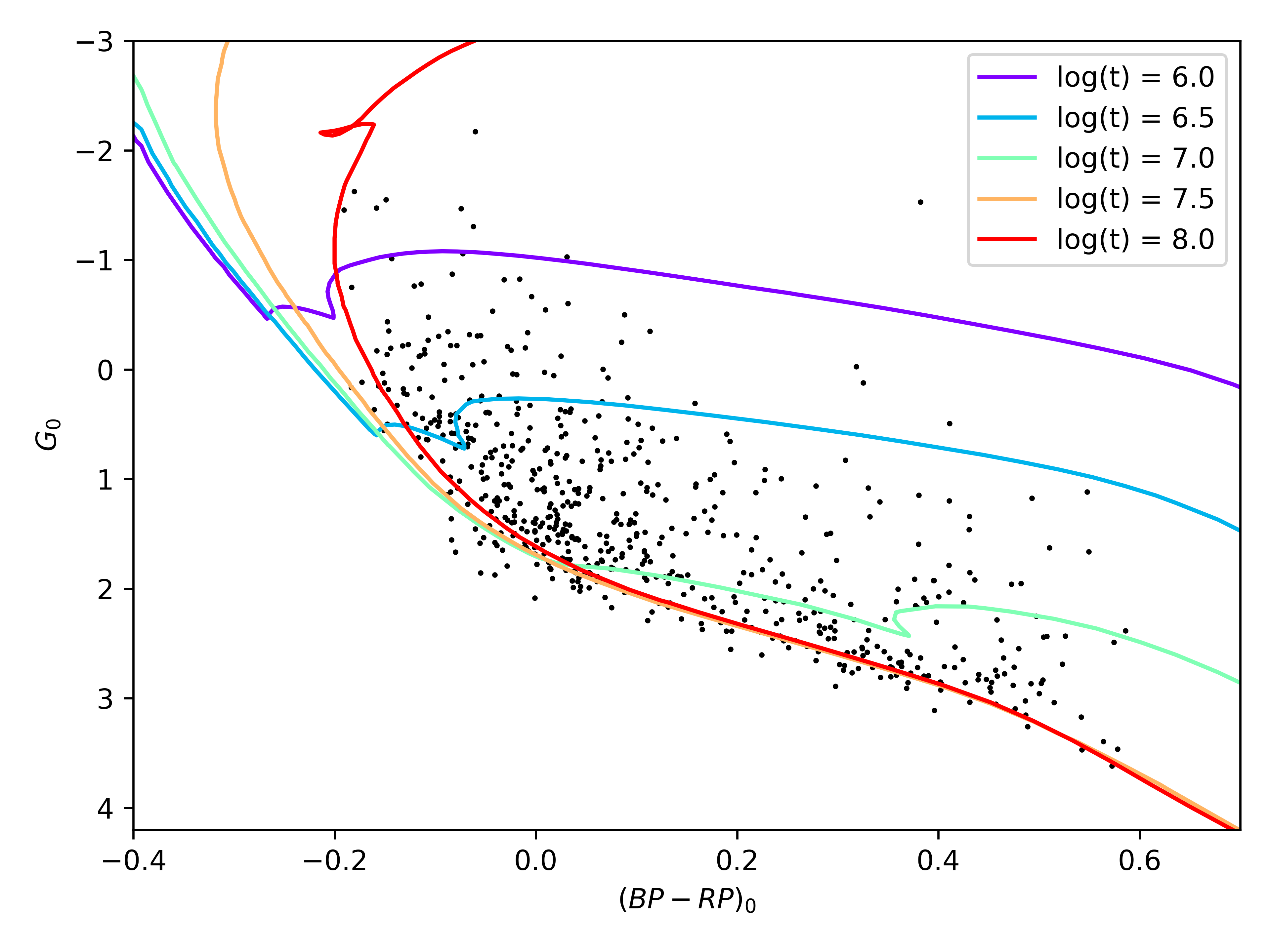}
    \caption{CMD 
 of the CP stars detected from spectroscopy.}
    \label{fig:cmd_spec}
\end{figure}
\unskip
\begin{figure}[H]
     
    \includegraphics[width=.9\columnwidth]{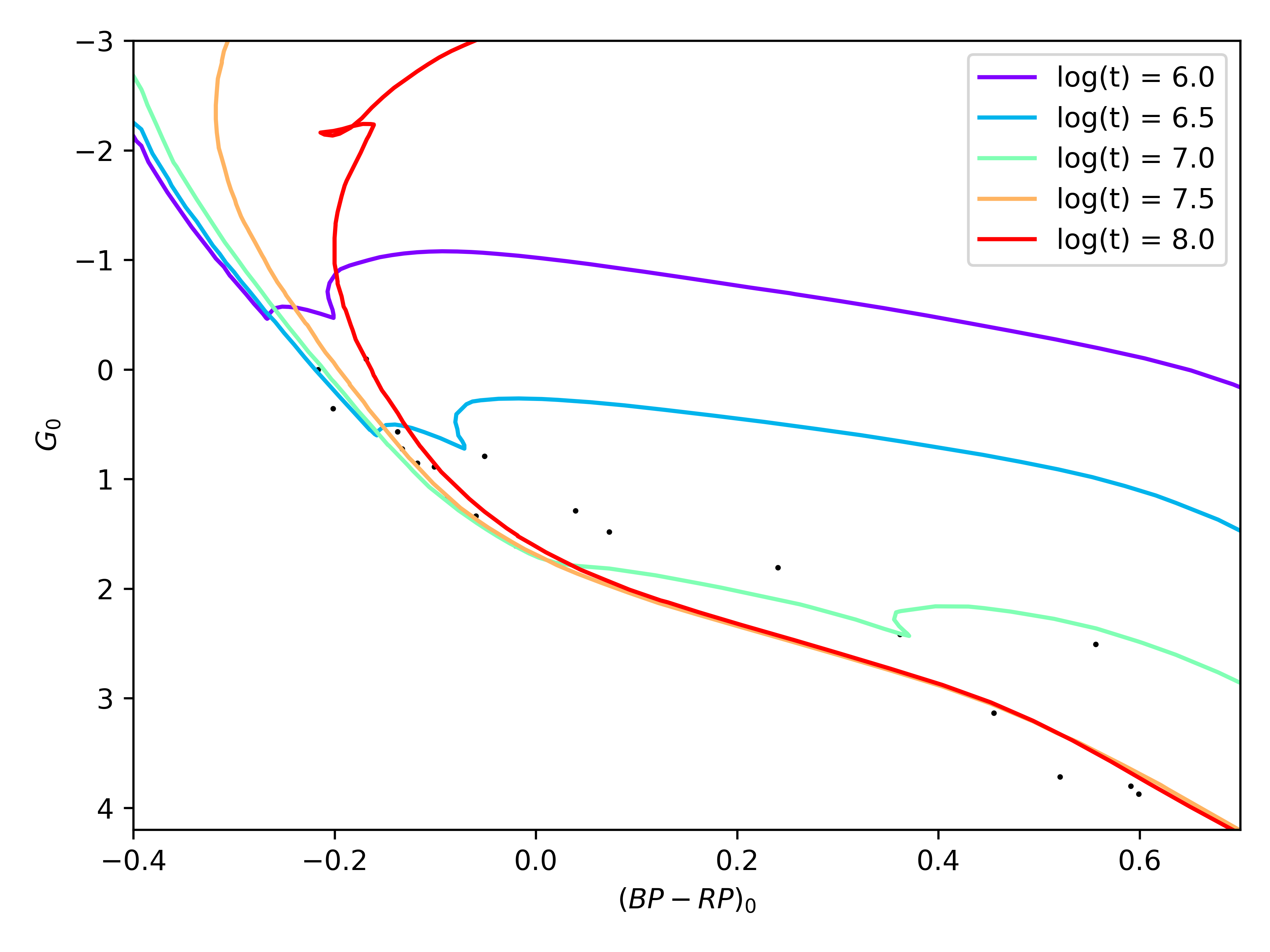}
    \caption{Same 
 as Figure \ref{fig:cmd_spec} but for the CP stars from light curves.}
    \label{fig:cmd_lcs}
\end{figure}

\startlandscape

\begin{adjustwidth}{+\extralength}{0cm}

\section[\appendixname~\thesection]{Spectroscopically Detected CP Stars}
 
\end{adjustwidth}


\begin{table}[H]
 \small
\caption{Stars that were classified as peculiar in at least one library from MKCLASS. The columns are (1) Gaia DR3 ID, (2) LAMOST ID, (3) right ascension $\alpha$ in degrees, (4)~declination $\delta$ in degrees, (5) parallax $\varpi$ and parallax error $\sigma_\varpi$, (6) open cluster/association the star is in, (7) classification from \texttt{libnor36}, (8) classification from \texttt{libr18}, and (9)~classification from \texttt{libr18\_225}. Only the first 30 entries are shown.}
 
\begin{tabularx}{\textwidth}{ccccccccc}
\toprule
\textbf{Source} & \textbf{LAMOST ID} & \boldmath{$\alpha$} & \boldmath{$\delta$} & \boldmath{$\varpi \pm \sigma_\varpi$} & \textbf{System} & \textbf{libnor36} 
 & \textbf{libr18} 
 &  \textbf{libr18\_225}
\\
 
\midrule
422486316486081280 
 &  J000940.81+575109.5 
& 2.42 & 57.853 & 0.288 $\pm$ 0.021 & UBC 1195 & B9 V & B9 V & kA0hA3mG8 \\
422850976388400896 & J001115.46+575859.3 & 2.814 & 57.983 & 0.295 $\pm$ 0.013 & UBC 1195 & kB9hF0mG8  Eu & F0 mA3 IV-V metal-weak & kB9.5hF0mG8  Eu \\
419025466196922752 & J002705.69+524904.1 & 6.774 & 52.818 & 0.566 $\pm$ 0.019 & Theia 2936 & kA8hA8mF4 & kA7hA8mF5 & kA7hA8mF4 \\
423487627981131648 & J010004.63+552553.8 & 15.019 & 55.432 & 0.767 $\pm$ 0.015 & AT 21 & A6 IV-V  (Sr) & A7 V & A7 V \\
423486318006267008 & J010036.07+552153.7 & 15.15 & 55.365 & 0.835 $\pm$ 0.012 & AT 21 & kA2hA3mA6  Si & A1 IV-V  SrSi & kA2hA6mA7  SrSi \\
522537579650553856 & J010816.40+613405.5 & 17.068 & 61.568 & 0.854 $\pm$ 0.023 & NGC 381 & F5 V (Sr) & F5 V (Sr) & F5 V (Sr) \\
426344193550526080 & J011006.42+602835.7 & 17.527 & 60.477 & 0.361 $\pm$ 0.018 & UBC 413 & kB9.5hF0mA5  Eu & kB9.5hA9mA7  Eu & kA0hA9mA7  Eu \\
410923920922247168 & J011920.73+551700.2 & 19.836 & 55.283 & 1.662 $\pm$ 0.015 & Theia 714 & A5 III  Sr & kA6hA9mF0 & A9 V \\
510674433300466816 & J012534.95+610827.0 & 21.396 & 61.141 & 0.383 $\pm$ 0.016 & COIN-Gaia 31 & kB9.5hF0mK0  Eu & B9.5 III & kA0hF0mK0  Eu \\
506120805889910272 & J015402.86+575513.7 & 28.512 & 57.92 & 0.419 $\pm$ 0.017 & UBC 1577 & kA1hF5mK2 & A0 III-IV & A8 mA1 V metal-weak \\
504941442233879296 & J015618.23+565650.4 & 29.076 & 56.947 & 0.455 $\pm$ 0.021 & UBC 1577 & B9 III-IV  Si & B9.5 III-IV  Si & F0 mA0 V metal-weak \\
504940239643050752 & J015627.82+565457.6 & 29.116 & 56.916 & 0.41 $\pm$ 0.019 & UBC 1577 & kA2hF5mK0 & A1 V & kA3hA3mK0 \\
504942473026002432 & J015707.89+565841.0 & 29.283 & 56.978 & 0.447 $\pm$ 0.017 & UBC 1577 & A2 V & A2 V & A2 V  Eu \\
504868904521760768 & J015803.44+563443.4 & 29.514 & 56.579 & 0.452 $\pm$ 0.014 & UBC 1577 & A1 V & kA1hA2mA4 & kA2hA3mG2 \\
504535791156850176 & J015815.54+553501.6 & 29.565 & 55.584 & 0.763 $\pm$ 0.017 & NGC 744 & A1 V & A0 V & kA1hA3mK0 \\
504523765248458624 & J015831.44+553159.6 & 29.631 & 55.533 & 0.725 $\pm$ 0.02 & NGC 744 & kA6hA8mF0 & kA6hA8mF1 & kA6hA8mF0 \\
505262671425427712 & J015954.67+570441.4 & 29.978 & 57.078 & 0.448 $\pm$ 0.023 & UBC 1577 & A1 V & DZ & kA2hA3mK0 \\
505103272298128896 & J020240.75+571130.1 & 30.67 & 57.192 & 0.406 $\pm$ 0.019 & UBC 1577 & kB9.5hF4mK2 & kB9.5hA1mA2 & kB9.5hF1mK0 \\
505086371610011904 & J020337.90+570205.8 & 30.908 & 57.035 & 0.427 $\pm$ 0.016 & UBC 1577 & A3 mA0 V Lam Boo & A0 IV-V & A3 mA0 V Lam Boo \\
505036412548787072 & J020523.60+565926.4 & 31.348 & 56.991 & 0.412 $\pm$ 0.015 & UBC 1577 & F0 mB8 V Lam Boo & F0 mA3 IV-V metal-weak & kB9hF0mG2  Eu \\
458378499188409856 & J021851.84+571255.0 & 34.716 & 57.215 & 0.38 $\pm$ 0.014 & NGC 869 & kB9hF0mA1  Eu & F0 V Fe-0.9 & F0 mA2 V metal-weak \\
458455121404670720 & J022212.78+571021.1 & 35.553 & 57.173 & 0.404 $\pm$ 0.02 & NGC 884 & A0 IV & A0 IV-V & kA0hA3mK1 \\
458406742891002496 & J022246.66+570448.8 & 35.694 & 57.08 & 0.383 $\pm$ 0.019 & NGC 884 & kB9hA8mK0  Eu & B9.5 IV & kA0hA8mK0  Eu \\
450639616650222720 & J023611.58+502045.3 & 39.048 & 50.346 & 1.377 $\pm$ 0.019 & COIN-Gaia 8 & A1 V & A3 mA0 metal weak & kA2hA3mK0 \\
450567632995957888 & J023759.19+495427.9 & 39.497 & 49.908 & 1.385 $\pm$ 0.02 & COIN-Gaia 8 & kA2hF5mK0 & A2 V & kA3hF5mK0 \\
436028657602103168 & J030153.84+474751.2 & 45.474 & 47.798 & 0.414 $\pm$ 0.019 & UBC 1246 & kB9.5hF1mG2  Eu & B9 III & kA0hF1mG2  Eu \\
435058750911681280 & J031436.47+471151.0 & 48.652 & 47.197 & 0.292 $\pm$ 0.018 & NGC 1245 & F0 V  Eu & A9 V  (Sr) & A9 V  Eu \\
435065038743702528 & J031439.44+471619.9 & 48.664 & 47.272 & 0.316 $\pm$ 0.02 & NGC 1245 & ? 
 & A9 V & F0 V  Eu \\
435064660786649088 & J031445.87+471246.9 & 48.691 & 47.213 & 0.292 $\pm$ 0.017 & NGC 1245 & A8 IV-Vn & kA7hA9mF1 & kA7hA8mF0 \\
435064798225554688 & J031449.77+471438.3 & 48.707 & 47.244 & 0.316 $\pm$ 0.016 & NGC 1245 & A4 IV  Eu & A7 V  Eu & A6 IV  Eu \\
... & ... & ... & ... & ... & ... & ... & ... & ...\\
\bottomrule
\end{tabularx}
\label{tab:spec_detections}
\end{table}

\finishlandscape
\section[\appendixname~\thesection]{Detection via \boldmath{$\Delta a$} Photometry}

\begin{table}[H]
 \small
\caption{CP stars detected via $\Delta a$ photometry with their basic astrometric properties. The columns are as follows: (1) \textit{Source}: the \textit{Gaia} DR3 ID, (2) $\alpha$: the right ascension in degrees, (3) $\beta$: the declination in degrees, (4) $\varpi$: the parallax and its error $\sigma_\varpi$ in mas, (5) \textit{System}: The cluster or association the star belongs to, (6) The colour $(g_1-y)$ from the synthetic photometry, (7) $\Delta a$: the $\Delta a$ value from the photometry. The stars are sorted by their right ascension $\alpha$. Again, the first 30 entries are shown.}

\begin{adjustwidth}{-\extralength}{0cm}

\begin{tabularx}{\fulllength}{cCCCCCC}
\toprule
\textbf{Source} & \boldmath{$\alpha$} & \boldmath{$\delta$} & \boldmath{$\varpi \pm \sigma_\varpi$} & \textbf{System} & \boldmath{$(g_1-y)$} & \boldmath{$\Delta a$} \\
 
\midrule
5436414262899903872  
 & 144.758 & $-$36.896 & 0.227 $\pm$ 0.044 & HSC 2138 & 15.122 & 1.707 \\
2930678291716664192 & 110.647 & $-$18.713 & 0.251 $\pm$ 0.042 & FSR 1252 & 11.584 & 2.208 \\
461829660022360832 & 50.986 & 59.24 & 0.953 $\pm$ 0.015 & Theia 1722 & 10.614 & 1.777 \\
5960713736908382336 & 265.155 & $-$40.215 & 0.673 $\pm$ 0.039 & Trumpler 29 & 11.36 & 1.423 \\
5881393391868603264 & 226.247 & $-$55.643 & 0.523 $\pm$ 0.014 & NGC 5823 & 9.197 & 2.32 \\
5881395625251644288 & 226.379 & $-$55.536 & 0.566 $\pm$ 0.013 & NGC 5823 & 9.957 & 2.022 \\
5881396037568526336 & 226.393 & $-$55.515 & 0.526 $\pm$ 0.013 & NGC 5823 & 9.775 & 1.319 \\
5881345799292540288 & 226.371 & $-$55.671 & 0.542 $\pm$ 0.013 & NGC 5823 & 9.324 & 1.371 \\
5881345837987362304 & 226.387 & $-$55.669 & 0.593 $\pm$ 0.013 & NGC 5823 & 9.113 & 1.46 \\
5881348414936867584 & 226.425 & $-$55.581 & 0.526 $\pm$ 0.017 & NGC 5823 & 9.096 & 1.444 \\
200085728707316096 & 73.536 & 40.35 & 0.932 $\pm$ 0.024 & HSC 1284 & 12.231 & 1.547 \\
4145604723045931136 & 273.091 & $-$16.472 & 0.539 $\pm$ 0.021 & OC 0025 & 9.862 & 1.503 \\
5716196751916693760 & 114.074 & $-$20.58 & 0.362 $\pm$ 0.029 & NGC 2421 & 11.456 & 1.89 \\
5716196958080774144 & 114.015 & $-$20.578 & 0.344 $\pm$ 0.043 & NGC 2421 & 11.495 & 1.698 \\
5715446026000379264 & 114.086 & $-$20.613 & 0.405 $\pm$ 0.038 & NGC 2421 & 10.768 & 1.602 \\
436077895107052928 & 45.559 & 47.929 & 0.394 $\pm$ 0.038 & UBC 1246 & 11.529 & 1.429 \\
436088684064785664 & 45.472 & 48.101 & 0.275 $\pm$ 0.043 & UBC 1246 & 11.257 & 1.591 \\
1870942218639565696 & 312.984 & 37.477 & 0.797 $\pm$ 0.017 & Roslund 7 & 12.453 & 1.41 \\
6071466104392664064 & 186.029 & $-$58.125 & 0.389 $\pm$ 0.024 & NGC 4337 & 11.567 & 1.583 \\
5520069849191410304 & 124.238 & $-$44.843 & 0.055 $\pm$ 0.037 & OC 0473 & 10.916 & 1.468 \\
$\dots$ & $\dots$ & $\dots$ & $\dots$ & $\dots$ & $\dots$ & $\dots$ \\
\bottomrule
\end{tabularx}
\label{tab:delta_detec}
\end{adjustwidth}
\end{table} 

\section[\appendixname~\thesection]{CP Stars from Photometric Time Series}

\begin{table}[H]
\small
\caption{Stars 
 with light curves that show CP properties. The columns are (1) \textit{Gaia} DR3 ID, (2) right ascension $\alpha$ in degrees, (3) declination $\delta$ in degrees, (4) parallax $\varpi$ and parallax error $\sigma_\varpi$, (5) the name of the association the star is in, designation from \cite{2021ApJ...917...23K}, (6) variable star type according to the VSX, (7)~the FAP of the period, (8) the period of the variability.}

\begin{adjustwidth}{-\extralength}{0cm}

\begin{tabularx}{\fulllength}{ccccccccc}

\toprule
 
\textbf{Source} & \textbf{TIC} & \boldmath{$\alpha$} & \boldmath{$\delta$} & \boldmath{$\varpi \pm \sigma_\varpi$} & \textbf{System} & \textbf{Type} & \boldmath{$log(FAP)$} & \boldmath{$P(d)$}\\
\midrule
 
3314527696567187584 
 & 17560109  
 & 67.283 & 18.455 & $4.531 \pm 0.020$ & Taur Ori 4 & ACV & $-\infty$ & 1.017 \\
3238478566084270208 & 397061497 & 75.275 & 3.717 & $3.479 \pm 0.048$ & Taur Ori 4 & SXARI & $-\infty$ & 2.42 \\
3207058750010997632 & 306345958 & 78.520 & $-$7.917 & $3.728 \pm 0.022$ & Greater Ori & ACV & $-$128.252 & 4.028 \\
3213882697128481792 & 4070682 & 80.103 & $-$3.510 & $3.181 \pm 0.020$ & Greater Ori & SXARI & $-\infty$ & 1.043 \\
3346259567786672384 & 247675468 & 86.855 & 12.648 & $3.351 \pm 0.037$ & Taur Ori 4 & ACV & $-\infty$ & 0.952 \\
3346656590266363648 & 247763365 & 87.357 & 13.922 & $3.654 \pm 0.030$ & Taur Ori 4 & ACV & $-\infty$ & 0.382 \\
3330219445485837184 & 436332628 & 91.617 & 10.750 & $5.555 \pm 0.041$ & Taur Ori 2 & SXARI & $-\infty$ & 0.362 \\
3104244792087711360 & 42884620 & 97.045 & $-$4.899 & $3.184 \pm 0.045$ & Mon SW & SXARI & $-\infty$ & 1.173 \\
5514764057046837248 & 388354822 & 124.907 & $-$50.382 & $3.128 \pm 0.016$ & Vela CG4 & ACV & $-\infty$ & 1.439 \\
5515989290962253184 & 82893704 & 125.065 & $-$47.969 & $3.400 \pm 0.020$ & Vela CG4 & ACV & $-$300.311 & 2.501 \\
5322386214086050176 & 89943694 & 127.570 & $-$50.941 & $3.082 \pm 0.011$ & Vela CG4 & ACV & $-\infty$ & 1.463 \\
5424236729248685824 & 77105310 & 141.327 & $-$43.624 & $5.625 \pm 0.012$ & IC 2391 & ACV & $-\infty$ & 1.471 \\
5232258082033470720 & 397732137 & 159.330 & $-$69.363 & $4.652 \pm 0.022$ & Cha & ACV & $-$222.997 & 1.584 \\
5239734932919731456 & 461125565 & 159.573 & $-$65.042 & $6.606 \pm 0.030$ & IC 2391 & SXARI & $-\infty$ & 0.549 \\
5239736204230183680 & 464907721 & 159.878 & $-$65.083 & $6.652 \pm 0.012$ & IC 2391 & ACV & $-\infty$ & 0.86 \\
4523888885382756224 & 50325659 & 278.215 & 17.303 & $3.179 \pm 0.019$ & Cerberus & SXARI & $-\infty$ & 2.99 \\
2262275415015759872 & 229793060 & 284.719 & 69.531 & $5.115 \pm 0.037$ & Cep Far North & ACV & $-\infty$ & 1.126 \\
2254485134616928128 & 229954300 & 288.681 & 65.273 & $4.858 \pm 0.031$ & Cep Far North & SXARI & $-\infty$ & 1.306 \\
2301907242919520640 & 236003103 & 305.141 & 82.848 & $4.238 \pm 0.022$ & Cep Far North & ACV & $-$306.611 & 1.323 \\
2190139274622512128 & 429331590 & 313.699 & 57.611 & $3.012 \pm 0.021$ & Cep Cyg & ACV & $-$220.353 & 1.417 \\
\bottomrule
\end{tabularx}
\end{adjustwidth}
\label{tab:lc_detec}
\end{table} 

\section[\appendixname~\thesection]{CP Candidate Stars with IR Excess}

\begin{table}[H]
 
\caption{List of the CP Stars and candidates that show IR excess.\label{tab: ir_excess_table}}

\begin{tabularx}{\textwidth}{CC}
\toprule
\textbf{\textit{Gaia} DR3 ID} & \textbf{Filter Where Excess Starts} \\
 
\midrule
182621906350208384  & WISE/WISE.W4 \\
183323227266520704 & WISE/WISE.W4 \\
183337177320364928 & WISE/WISE.W1 \\
1871158238309294208 & WISE/WISE.W3 \\
187964540726357632 & WISE/WISE.W1 \\
2007069351654097664 & WISE/WISE.W1 \\
2070344734000894592 & WISE/WISE.W3 \\
2075877648313310848 & WISE/WISE.W1 \\
2075877678366383744 & WISE/WISE.W4 \\
2076066351994581120 & WISE/WISE.W4 \\
2085246793048249344 & WISE/WISE.W3 \\
216590390373910912 & Spitzer/IRAC.I4 \\
216617641943232000 & Spitzer/MIPS.24mu \\
251761469543936512 & WISE/WISE.W3 \\
253455576447809280 & WISE/WISE.W4 \\
3017409624342379520 & WISE/WISE.W3 \\
3019790754200699136 & WISE/WISE.W1 \\
3019875485316186752 & WISE/WISE.W1 \\
3125739694655670912 & UKIRT/UKIDSS.K \\
3209504544905800576 & WISE/WISE.W4 \\
3210561587897894528 & WISE/WISE.W4 \\
3304135250100217600 & WISE/WISE.W2 \\
3330884473923251968 & WISE/WISE.W4 \\
3342955019950652928 & WISE/WISE.W4 \\
3355226737949262336 & WISE/WISE.W4 \\
3355241409558154368 & WISE/WISE.W4 \\
3370708239624841344 & WISE/WISE.W3 \\
3372437530897057920 & WISE/WISE.W4 \\
3372466874113370240 & WISE/WISE.W4 \\
3373851365410096896 & WISE/WISE.W3 \\
3373860500801601408 & WISE/WISE.W3 \\
3377260401211784448 & WISE/WISE.W3 \\
3377541841828979456 & WISE/WISE.W4 \\
3383565585001209728 & WISE/WISE.W3 \\
3383771812150964096 & WISE/WISE.W3 \\
3394744216639326208 & WISE/WISE.W3 \\
3444662212044370944 & WISE/WISE.W3 \\
3445205168927696512 & WISE/WISE.W4 \\
3455831880089411072 & WISE/WISE.W3 \\
3455880567838460544 & WISE/WISE.W3 \\
435065038743702528 & WISE/WISE.W1 \\
4469325071096213504 & WISE/WISE.W3 \\
458406742891002496 & WISE/WISE.W1 \\
473267192293775360 & WISE/WISE.W1 \\
48058592395097856 & WISE/WISE.W3 \\
1824774481337240576 & WISE/WISE.W1 \\
1981253265315002880 & WISE/WISE.W3 \\
2015462885981225984 & WISE/WISE.W3 \\
2028730486687781248 & WISE/WISE.W4 \\
2055676974020815488 & WISE/WISE.W1 \\
2058950666823409152 & WISE/WISE.W4 \\

\bottomrule
\end{tabularx} 
 
\end{table}

\begin{table}[H]\ContinuedFloat
 
\caption{{\em Cont.}}
\begin{tabularx}{\textwidth}{CC}
\toprule
\textbf{\textit{Gaia} DR3 ID} & \textbf{Filter Where Excess Starts} \\
 
\midrule

2164328273542938880 & UKIRT/UKIDSS.K \\
2178496541693778688 & WISE/WISE.W3 \\

2930678291716664192 & WISE/WISE.W1 \\
2931333566285206912 & WISE/WISE.W1 \\
2931478907978887424 & WISE/WISE.W1 \\
3051142842935888512 & WISE/WISE.W4 \\
3124537309971018752 & WISE/WISE.W1 \\
3125525015011266304 & UKIRT/UKIDSS.K \\
3126379163747692672 & WISE/WISE.W3 \\
3126387135207429376 & WISE/WISE.W3 \\
3126748977611495936 & WISE/WISE.W1 \\
3128813203316883456 & WISE/WISE.W3 \\
3129234045689550976 & WISE/WISE.W1 \\
3324289470039249408 & WISE/WISE.W1 \\
3352222215048693504 & WISE/WISE.W3 \\
3355188117603280640 & WISE/WISE.W1 \\
4064497976508146688 & WISE/WISE.W2 \\
428891315316060160 & WISE/WISE.W1 \\
435067993681192320 & UKIRT/UKIDSS.K \\
464638293759397632 & WISE/WISE.W1 \\
465827037629837696 & WISE/WISE.W4 \\
5254330709292306560 & WISE/WISE.W1 \\
5254332770876859008 & WISE/WISE.W4 \\
5310688514207181824 & WISE/WISE.W1 \\
5333062854296532480 & WISE/WISE.W4 \\
5335935942551982080 & WISE/WISE.W2 \\
5337166162957067648 & WISE/WISE.W1 \\
5337376891236690560 & WISE/WISE.W3 \\
5338495403810748416 & AKARI/FIS.WIDE-S \\
5338502202699086080 & WISE/WISE.W1 \\
5339377246496587392 & WISE/WISE.W1 \\
5339411846756340736 & WISE/WISE.W1 \\
5339894291828433024 & WISE/WISE.W1 \\
5339935695306646016 & WISE/WISE.W3 \\
5339942189297580416 & WISE/WISE.W3 \\
5351221941647302912 & WISE/WISE.W1 \\
5351241973377009280 & WISE/WISE.W1 \\
5352004244161147136 & WISE/WISE.W1 \\
546642482291623168 & WISE/WISE.W3 \\
5518730442165840768 & WISE/WISE.W1 \\
5520118060206928896 & WISE/WISE.W1 \\
5547527132743054336 & WISE/WISE.W1 \\
5547599769226387840 & WISE/WISE.W3 \\
5547978963304967296 & WISE/WISE.W1 \\
5619904272346555520 & WISE/WISE.W3 \\
5715306662901791232 & WISE/WISE.W2 \\
5865056126594144128 & WISE/WISE.W2 \\
6054029567979605632 & WISE/WISE.W1 \\
6054273556477806976 & WISE/WISE.W1 \\
6054329940813382016 & WISE/WISE.W2 \\
6054868770221377920 & Spitzer/IRAC.I4 \\
6054918488763944832 & WISE/WISE.W2 \\
6056528822281714816 & WISE/WISE.W1 \\
\bottomrule
\end{tabularx}
\end{table}

\startlandscape

\begin{adjustwidth}{+\extralength}{0cm}
\section[\appendixname~\thesection]{CP Candidate Stars with H\boldmath{$\alpha$} Emission}
 
\end{adjustwidth}


\begin{table}[H]
\caption{CP candidates that show emission lines. The columns are (1) \textit{Gaia} DR3 ID, (2) LAMOST ID, (3) right ascension $\alpha$ in degrees, (4) declination $\delta$ in degrees, (5)~parallax $\varpi$ and parallax error $\sigma_\varpi$, (6) the colour $(g1_-y)$ from the synthetic photometry, (7) the H$\alpha$ index, (8)--(10) the spectral types derived from MKCLASS.}

\begin{tabularx}{\textwidth}{cccccccccC}
 
\toprule
\textbf{\textit{Gaia} DR3 ID} & \textbf{LAMOST} & \boldmath{$\alpha$} & \boldmath{$\delta$} & \boldmath{$\varpi \pm \sigma_\varpi$} & \boldmath{$(g_1-y)$} & \textbf{H\boldmath{$\alpha$}} & \texttt{\textbf{libr18}} & \texttt{\textbf{libnor36}} & \texttt{\textbf{libr18\_225}} \\
 
\midrule
183576114939066752 & J052117.64+345221.5 & 80.324 & 34.873 & 0.325 $\pm$ 0.03 & 9.529 & 1.903 & kB9hF0mK0  Eu & B9.5 III & kB9.5hF0mG2  Eu \\
3131334967590689536 & J063207.38+045455.8 & 98.031 & 4.916 & 0.649 $\pm$ 0.017 & 11.703 & 1.729 & F0 mB5 V Lam Boo & B7 V & kB8hF0mK0  Eu \\
182636268718056576 & J052815.47+342525.6 & 82.064 & 34.424 & 0.489 $\pm$ 0.02 & 9.835 & 1.486 & B4 IV-V & B4 V & kB7hF2mF9 \\
2059202695503955584 & J200754.27+361314.1 & 301.976 & 36.221 & 0.316 $\pm$ 0.022 & 9.638 & 1.406 & kA2hA3mA6 & A1 IV & kA2hA3mA7 \\
2007069351654097664 & J225100.01+570219.6 & 342.75 & 57.039 & 0.297 $\pm$ 0.011 & 9.616 & 1.326 & kB9.5hF0mG2  Eu & B9 II-III & kA0hF0mG8  Eu \\
182694783351830912 & J052405.96+340124.2 & 81.025 & 34.023 & 0.411 $\pm$ 0.016 & 9.041 & 1.668 & B9 III & B9.5 II-III & kA0hF1mG2 \\
3343027622077611392 & J055508.01+115959.2 & 88.783 & 12 & 0.639 $\pm$ 0.025 & 9.763 & 1.851 & F0 mB8 V Lam Boo & B9 IV & F0 mA1 V metal-weak \\
182635620182511744 & J052805.13+342130.3 & 82.021 & 34.358 & 0.461 $\pm$ 0.015 & 9.466 & 1.871 & B6 V & B7 V & kB9hF1mK0  Eu \\
1979250470521255424 & J214103.58+503109.8 & 325.265 & 50.519 & 0.241 $\pm$ 0.019 & 9.172 & 1.691 & B9.5 II-III & A0 II-III & kA0hF0mA5  Eu \\
3131335040608985216 & J063213.98+045514.2 & 98.058 & 4.921 & 0.671 $\pm$ 0.017 & 12.859 & 1.57 & B8 IV  Si & emission-line ? 
 & F1 V Fe-0.8 \\
182659976937528320 & J052819.83+342720.1 & 82.083 & 34.456 & 0.803 $\pm$ 0.016 & 10.12 & 1.526 & kA2hF5mK0 & A2 V & kA2hA3mK0  Eu \\
3372466255638764928 & J062601.84+194004.7 & 96.508 & 19.668 & 0.275 $\pm$ 0.018 & 9.576 & 1.723 & kB9hF0mK0  Eu & B9 III & kB9.5hF0mK0  Eu \\
\bottomrule
\end{tabularx}
\label{tab:emission_stars}
 
\end{table}

\finishlandscape

\begin{adjustwidth}{-\extralength}{0cm}

\reftitle{References}


\PublishersNote{}
\end{adjustwidth}

\end{document}